\documentclass[a4paper,11pt]{article}
\usepackage{jheppub} 
\usepackage{lineno}
\usepackage{float}
\usepackage{slashed}
\usepackage{xfrac,bigints}
\usepackage{physics}
\usepackage{subfigure}
\usepackage{booktabs} 
\usepackage{mathtools}

\title{Geometry Induced Chiral Transport and Entanglement in $AdS_2$ Background}
\author[a,b]{Kazuki Ikeda}
\emailAdd{kazuki.ikeda@umb.edu}
\affiliation[a]{Department of Physics, University of Massachusetts Boston}
\affiliation[b]{Center for Nuclear Theory, Department of Physics and Astronomy, Stony Brook University}
\author[c]{Yaron Oz}
\affiliation[c]{School of Physics and Astronomy, Tel-Aviv University,
Tel-Aviv 69978, Israel}
\emailAdd{yaronoz@tauex.tau.ac.il}
\abstract{
We study the real-time chiral dynamics of Dirac fermions in AdS$_2$ and AdS$_2$ black hole backgrounds. The spacetime curvature generates a spin connection term, acting as an effective magnetic field and a position-dependent chiral chemical potential. This leads to strongly asymmetric wave propagation, confined within an inhomogeneous Lieb-Robinson cone. The front velocities decrease with increasing fermion mass and horizon radius. The entanglement entropy grows inside the causal cone, and it saturates due to screening/dephasing in the finite inhomogeneous chain.
 In dipole-dipole collision, the central bipartite entropy rises when the inward Lieb-Robinson fronts intersect, forming a bright ridge in the local entanglement profile. Charge and current correlators peak at the front arrival, providing a real-time diagnostic of chiral transport. These results establish a causality-respecting framework, linking curvature and horizons to transport and entanglement in (1+1)-dimensional fermionic matter.
}

\begin{document}
\maketitle
\flushbottom

\section{Introduction}

The interplay between spacetime geometry, quantum transport, and entanglement dynamics lies at the heart of modern high energy and quantum many body physics. Gravitational backgrounds provide natural arenas to study causal propagation, thermalization, and information flow. Within the AdS/CFT correspondence \cite{Maldacena1998,GubserKlebanovPolyakov1998,Witten1998AdS} (for a review see \cite{AharonyGubserMaldacenaOoguriOz2000}), such setups offer a controlled bridge between curved spacetime dynamics and lower-dimensional quantum field theories, motivating explicit real-time studies of quantum matter in curved geometries.

In this work, we investigate the real-time chiral dynamics of Dirac fermions in AdS$_2$ and AdS$_2$ black hole backgrounds, focusing on how curvature and horizons imprint geometric signatures on transport and entanglement. The spin connection in curved spacetime acts as an effective magnetic field, and induces a position-dependent chiral chemical potential, even in the absence of gauge interactions. This curvature-induced bias breaks left-right symmetry, and generates asymmetric propagation of fermionic excitations, providing a purely geometric origin for chiral transport.

We construct a qubit Hamiltonian for the Dirac field using staggered fermions ~\cite{Kogut:1974ag,Susskind:1976jm} and a Jordan-Wigner transformation \cite{Jordan:1928wi}, mapping the continuum theory to a one-dimensional spin chain with redshift-weighted couplings, and Dzyaloshinskii-Moriya type \cite{Dzyaloshinskii1958,Moriya1960} interactions arising from the spin connection. Real-time dynamics are simulated with matrix product states (for a review see \cite{Orus2014}), enabling direct access to charge transport, entanglement growth, and correlation functions in AdS$_2$ and AdS$_2$ black hole spacetimes.

The simulations reveal strongly left-right asymmetric wavefronts, confined within an inhomogeneous Lieb-Robinson (LR) cone \cite{Lieb:1972wy}, whose velocities decrease with increasing fermion mass or horizon radius. Entanglement growth occurs within this causal cone and eventually saturates due to screening and dephasing in the finite inhomogeneous chain. In dipole-dipole collision, the onset of central bipartite entropy coincides precisely with the causal meeting of the inward LR fronts, demonstrating that entanglement generation obeys a curvature-modified causality bound. Furthermore, charge-charge and current-current correlators peak at the front arrival, providing a real-time diagnostic of chiral propagation. Taken together, these results establish a causality-respecting framework for characterizing how curvature and horizons control transport and entanglement in (1+1)-dimensional fermionic matter. The approach connects gravitational physics with quantum information theory, and is directly compatible with programmable quantum simulators capable of engineering spatially varying couplings.

From the perspective of quantum many-body dynamics, placing a fermionic system in a curved metric allows us to study non-perturbative effects beyond solvable models: the geometry reshapes the local notion of time and distance and therefore translates, after discretization, into a spatially inhomogeneous Hamiltonian. In the present black hole setting this inhomogeneity appears in two distinct ways. First, the gravitational redshift produces position-dependent kinetic couplings, leading to a bent (inhomogeneous) causal fan. Second, the spin connection contributes an additional term that acts as a
position-dependent chiral bias (an effective $\mu_5$), generating left-right asymmetric real-time propagation without gauge interactions or externally imposed drives. The black hole horizon radius $r_h$ provides a tunable control parameter for how strongly this geometric inhomogeneity suppresses transport and entanglement.

Real-time tensor-network studies of (1+1)d fermions and lattice gauge theories in flat space typically involve homogeneous couplings, where the causal cone is straight and parity symmetric unless chirality is introduced by external fields/chemical potentials or interactions. Here we bridge curved-spacetime fermion dynamics with modern quantum many-body simulation by (i) deriving an explicit local qubit Hamiltonian whose couplings encode the redshift and spin connection and (ii) characterizing chiral transport and entanglement growth using quantitative causality diagnostics.

The paper is organized as follows. Section~\ref{sec:fermion} reviews Dirac fermions in two-dimensional curved spacetime and notations for the covariant derivative, spin connection, and conserved quantities. Section~\ref{sec:dynamics} specializes to AdS$_2$ and AdS$_2$ black hole backgrounds, deriving the continuum action/Hamiltonian and constructing the lattice (qubit) Hamiltonian via Jordan–Wigner transformation. It also defines the AdS–weighted charge and current operators used throughout the numerics. Sec.~\ref{sec:dynamics} presents the real‑time quench protocol and analyzes chiral dynamics: Sec.~\ref{chiral} discusses geometry‑induced chiral waves; Sec.~\ref{subsec:chiral-ent-and-charge} studies charge separation and entanglement growth following a localized dipole quench; Sec.~\ref{sec:LR} develops inhomogeneous LR bounds and arrival‑time diagnostics; and Sec.~\ref{sec:corr} introduces AdS‑weighted charge–charge and current–current correlators. Sec.~\ref{sec:scattering} examines dipole–dipole collision, first tracking charge propagation and interference of the two causal fans (Sec.~\ref{sec:scattering_charge}), then relating the growth of central bipartite entanglement to the collision of the inward LR branches (Sec.~\ref{sec:scattering_LR}). Section~\ref{sec:conclusion} concludes with a summary and outlook.

%\newpage
\section{\label{sec:fermion}Fermions on $AdS_2$ Black Hole}

In this section, we formulate the dynamics of Dirac fermions in two-dimensional curved spacetime, and derive the corresponding Hamiltonian for AdS$_2$ and AdS$_2$ black hole geometries. We first review the general framework for fermions in curved backgrounds. Specializing to the AdS$_2$ black hole metric, we obtain the continuum Lagrangian and Hamiltonian, identify the curvature-induced redshift and spin connection terms, and discretize the theory using staggered fermions. Applying the Jordan-Wigner transformation yields a qubit Hamiltonian, i.e. a spin-chain representation with position-dependent couplings, that encode the AdS geometry \cite{Ikeda:2025yqq}, and serve as the basis for the real-time simulations in later sections.

\subsection{Fermions on Two-Dimensional Curved Spaces}

%\subsection{Action and Hamiltonian}

We will work in two-dimensional curved space-time with metric $g_{\mu\nu}$, related to the flat Minkowski metric $\eta_{ab}=\mathrm{diag}(-,+)$ by the vielbein $e_\mu^a$ by:
\begin{equation}
g_{\mu\nu} = e_\mu^a e_\nu^b \eta_{ab} \ .
\label{VB}
\end{equation}
The gamma matrices in curved space $\gamma^{\mu}$ are related to the flat space gamma matrices $\gamma^a$ by:
\begin{equation}
\gamma^\mu = e_a^\mu \gamma^a \ ,
\end{equation}
where $e_\mu^a e_b^\mu = \delta^a_b$. 
The flat-frame gamma matrices satisfy the two-dimensional Clifford algebra:
\begin{equation}
\{\gamma^{a},\gamma^{b}\}= - 2\,\eta^{ab} \ ,
\end{equation}
and we will use the representation:
\begin{equation}
\gamma^0 = \sigma_z, \gamma^1 = i \sigma_y, \gamma^5=\gamma^0\gamma^1 = \sigma_x \ ,
\label{gnotation}
\end{equation}
and $(\gamma^0)^{\dagger} = \gamma^0,  (\gamma^1)^{\dagger} = -\gamma^1, 
(\gamma^5)^{\dagger} = \gamma^5$.

We consider a massive Dirac fermion, whose action reads:
\begin{equation}
S = \int d^2x \, \sqrt{-g} \, \bar{\psi} \left( i \gamma^\mu D_\mu - m  \right) \psi \ ,
\end{equation}
where the adjoint spinor is defined as $\bar{\psi}= \psi^{\dagger}\gamma^0$, where $\gamma^0$ is the flat space gamma matrix, and $A = (A_t,0)$ is a time-like vector field. $D_{\mu}$ is the fermionic covariant derivative:
\begin{equation}
D_\mu \psi = \partial_\mu \psi - \frac{1}{4} \omega_\mu^{ab} \gamma_a \gamma_b \psi \ ,    \end{equation}
where the torsionless spin connection $\omega^a{}_b = \omega^a{}_{b\mu}dx^{\mu}$ satisfies the 
Cartan equation:
\begin{equation}
de^a + \omega^a{}_b\wedge e^b=0 \ ,  
\end{equation}
where $e^a = e^{a}_{\mu}dx^{\mu}$.

\subsection{Fermion on $AdS_2$ Black Hole}

\subsubsection{Action and Hamiltonian}
Consider Schwarzschild black hole solution in AdS$_2$ with radius $L$:
\begin{equation}
\mathrm{d}s^2 
\;=\;
-\,f(r)\,\mathrm{d}t^2 
\;+\;
\frac{1}{f(r)}\,\mathrm{d}r^2,
\quad
\text{where}
\quad
f(r) 
\;=\;
\frac{\,r^2 - r_h^2\,}{\,L^2\,} \ .
\label{AdSBH}
\end{equation}
Here $r_h$ is the horizon radius. In the units $16\pi G_{2}=1$ the mass of the black hole is $M=r_h^2$  and its temperature $T= \frac{r_h}{2\pi L^2}$. The zweibein read:
\begin{equation}
e^a_{\;\mu} \;=\;
\begin{pmatrix}
\sqrt{f(r)} & 0\\[4pt]
0 & \frac{1}{\sqrt{f(r)}}
\end{pmatrix},~~~
e_a^{\;\mu} \;=\;
\begin{pmatrix}
\frac{1}{\sqrt{f(r)}} & 0\\[4pt]
0 & \sqrt{f(r)} 
\end{pmatrix}.
\end{equation}
The nonzero spin connection is:
\begin{equation}
\omega_{t}^{01}
\;=\; \frac{r}{L^2} \ .
\end{equation}

When $r_h \to 0$, we have $f(r)
\;\longrightarrow\;
\frac{r^2}{\,L^2\,}$, and the metric becomes:
\begin{equation}
\mathrm{d}s^2 
\;\to\;
-\,
\frac{r^2}{\,L^2\,}\,\mathrm{d}t^2 
\;+\;
\frac{L^2}{\,r^2\,}\,\mathrm{d}r^2.
\end{equation}
Introducing a coordinate $z = -\tfrac{L^2}{r}$, we get the metric:
\begin{equation}
\mathrm{d}s^2
\;=\;
\frac{L^2}{\,z^2\,}
\Bigl(
-\,\mathrm{d}t^2 
\;+\;
\mathrm{d}z^2
\Bigr),
\end{equation}
which is the Poincar\'e AdS$_2$ form.

The Lagrangian density takes the form:\footnote{The operator
$\frac{i}{2}\{A(r),\partial_r\}\;=\;i\,A(r)\,\partial_r\;+\;\frac{i}{2}\,A'(r)$ is hermitian. Here   
$A(r) = \sqrt{f(r)}\,\sigma_x$.}
\begin{equation}
{\cal L} = \psi^{\dagger} \left[ i \frac{1}{\sqrt{f(r})} \partial_t + i \sqrt{f(r)} \sigma_x \partial_r +i \frac{r}{2L^2\sqrt{f(r)}} \sigma_x  - m \sigma_z + \mu \right] \psi \ .
\end{equation}
The conjugate momentum to $\psi$ is:
\begin{equation}
\Pi_{\psi} \equiv \frac{\partial{{\cal{L}}}}{\partial (\partial_t \psi)} =  \frac{i}{\sqrt{f(r)}} \psi^{\dagger} \ ,
\end{equation}
and the canonical anticommutation relations read:
\begin{equation}
\{\psi_\alpha(r,t),\psi_\beta^\dagger(r’,t)\}
=\delta_{\alpha\beta}\,\sqrt{f(r)}\,\delta(r-r’) \ ,
\label{ACR}
\end{equation}
where $\alpha$ and $\beta$ are the spinor indices.

We can flatten the inner product by a local rescaling: $\chi(r)=f(r)^{-1/4}\psi(r)$, hence:
\begin{equation}
\{\chi(r),\chi^\dagger(r’)\}=\delta(r-r’) \ . 
\label{ACRF}
\end{equation}
The Lagrangian density takes the form:
\begin{equation}
{\cal L} = \chi^{\dagger} \left[ i \partial_t + i f(r) \sigma_x \partial_r +i \frac{r}{L^2} \sigma_x  - \sqrt{f(r)}m \sigma_z + \sqrt{f(r)}\mu \right] \chi \ ,
\end{equation}
and the Hamiltonian reads:
\begin{equation}
    H = \int_{r_h}^{\infty} d r  \mathcal{H} = \int d r \chi^{\dagger} \left[
    - i f(r) \sigma_x \partial_r -i \frac{r}{L^2} \sigma_x  + \sqrt{f(r)}m \sigma_z - \sqrt{f(r)}\mu
    \right] \chi \ .
    \label{BHHF}
\end{equation}
In the limit $r_h \to 0$, the Hamiltonian reduces to the AdS$_2$ Hamiltonian with the Poincaré coordinates. Outside the horizon $r>r_h$, $\xi=\partial_t$ is a timelike Killing vector and the Hamiltonian is conserved and corresponds to the symmetry $\mathcal{L}_\xi$. The vector $\xi$ becomes null at the horizon and space-like inside the horizon, and while it remains a killing vector, we cannot use it to define a Hamiltonian flow. Inside the horizon $\partial_r$ is timelike and generates  evolution along infalling time-like geodesics. This, however, is not associated with a conserved energy measured at infinity, because the usual notion of energy is tied to asymptotic symmetries at the boundary. Hence, we are restricted to study the system's properties outside the horizon.

We will consider the weighted charge that reflects the underlying curved spacetime geometry:% that multiplies the chemical potential $\mu$ in the Hamiltonian (\ref{BHHF}):
\begin{equation}
Q_{weighted}=\int_{r_h}^\infty dr\,\sqrt{f(r)}\chi^\dagger\chi \ .  
\label{qweighted}
\end{equation}

The redshift factor is $\alpha(r)=\sqrt{-g_{tt}(r)} = \sqrt{f(r)}$, where $\alpha(r)$ is defined as:
\begin{equation}
\alpha(r)=\sqrt{\frac{r^2 - r_h^2}{L^2}} \ ,    
\label{redshift}
\end{equation}
which is the ratio of boundary time to near-horizon proper time $d\tau = \alpha dt$, and thus measures the gravitational redshift between the $AdS_2$ boundary where we define our field theory Hamiltonian, and the black hole throat where modes live. This is the gravitational redshift, that rescales all near-horizon energies and momenta by $\alpha$. Thus, a mode of frequency $\omega_{\rm throat}$ and momentum $k_{\rm throat}$ near the horizon is seen at the boundary with frequency
\begin{equation}
\omega_{\rm bdry} \;=\;\alpha\;\omega_{\rm throat},~~~k_{\rm bdry} \;=\;\alpha\;k_{\rm throat} \ .    
\end{equation}
At the horizon $\alpha=0$, hence a finite frequency at the horizon appears infinitely redshifted to the boundary — i.e., it has zero frequency from the boundary perspective. When going to the boundary we need to normalize the boundary clock so that the physical time is $t_{\rm bdy}$, hence at the boundary $\alpha \to 1$. The energy-momentum dispersion relation reads:
\begin{equation}
\varepsilon(k) = \alpha\sqrt{m^2 + (\alpha k)^2} \  . 
\label{dr}
\end{equation}
Note that the reason for the additional $\alpha$ factor in front of the wave number $k$ in (\ref{dr}) is that it is the momentum conjugate to $r$, and not to the proper spatial coordinate $\rho$, $d\rho = \frac{dr}{\alpha(r)}$.

\subsubsection{Qubit Hamiltonian}
To convert the Hamiltonian~\eqref{BHHF} into the lattice Hamiltonian, we use the staggered fermion $\chi_n$~\cite{Kogut:1974ag,Susskind:1976jm}
\begin{equation}
    \psi(t=0,z=na)=\frac{1}{\sqrt{a}}\begin{pmatrix}
        \chi_n\\
        \chi_{n+1}
    \end{pmatrix},
\end{equation}
and replace $\partial_z\psi(z)$ by a finite difference $\frac{\psi(z_{n+1})-\psi(z_n)}{a}$. The qubit-representation of the lattice Hamiltonian is obtained by Jordan-Wigner transformation~\cite{Jordan:1928wi}:
\begin{align}
\begin{split}
 \chi_n = \frac{X_n-i Y_n}{2}\prod_{i=1}^{n-1}(-i Z_i) \ ,
\quad \chi^\dag_n = \frac{X_n+i Y_n}{2}\prod_{i=1}^{n-1}(i Z_i) \ ,
\end{split}
\label{JW}
\end{align}
where $X_n,Y_n,Z_n$ are the Pauli matrices at the $n$-th site.

Upon the regularization, our Hamiltonian with mass is \cite{Ikeda:2025yqq} 
\begin{align}
    \begin{aligned}
        H=&\frac{1}{4a}\sum_{n=1}^{N-1}\alpha_n^2(X_nX_{n+1}+Y_nY_{n+1})+\frac{a}{8L^2}\sum_{n=1}^{N-1}n(X_nY_{n+1}-Y_nX_{n+1})\\
        &+\frac{m}{2}\sum_{n=1}^N(-1)^n\alpha_n(Z_n+1) \ , 
    \end{aligned}
    \label{eq:Ham_AdS}
\end{align}
where $\alpha_n = \sqrt{f(r_n)}$ is the redshift factor at site $n$:
\begin{equation}
\alpha_n = \sqrt{\frac{r_n^2 - r_h^2}{L^2}} \ .    
\label{redshiftL}
\end{equation}
The first term in the Hamiltonian is identical to the kinetic term in flat space. The subsequent term involves the spin connection, represented by $\frac{1}{2z}\psi^\dagger\sigma_x\psi=\frac{1}{2an}\bar{\psi}\gamma^1\psi$. The mass term is $L$-dependent. If we take the limit $L\to\infty$ while maintaining $\frac{L}{an}=1$, the Hamiltonian reduces to the flat space. 

The charge \eqref{qweighted} and current operators that reflect the AdS geometry are 
\begin{align}
\begin{aligned}
\label{QC}
 \text{Charge:}&\quad Q = \frac{1}{2a}\sum_{n=1}^N\alpha_n(Z_n+(-1)^n)\;, \\
\text{Current:}&\quad J\;=\;\frac{1}{4}\sum_{i=1}^{N-1}\alpha^2_i(X_iY_{i+1}-Y_iX_{i+1})  \ .
\end{aligned}
\end{align}

Correspondence between observables and Pauli operators are summarized below. (For the details, see \cite{Ikeda:2025yqq}) 
\begin{table}[H]
\begin{center}
\begin{tabular}{c|c|c}%\toprule
Dirac Fermion Bilinears& Staggered  & Pauli \\\hline
     $\overline{\psi}\psi$ & $\frac{(-1)^n}{a}\chi^\dagger_n\chi_n$ &  $\frac{(-1)^n}{2a}(Z_n+1)$ \\
     $\overline{\psi}\gamma_0\psi$ & $\frac{1}{a}\chi^\dagger_n\chi_n$ &  $\frac{1}{2a}(Z_n+(-1)^n)$ \\
     $\overline{\psi}\gamma_1\psi$ & $\frac{1}{2a}(\chi^\dagger_n\chi_{n+1}+\chi^\dagger_{n+1}\chi_{n})$ &  $\frac{1}{4a}(X_nY_{n+1}-Y_nX_{n+1})$ \\
    $\overline{\psi}\gamma_5\psi$ & $\frac{(-1)^n}{2a}(\chi^\dagger_n\chi_{n+1}-\chi^\dagger_{n+1}\chi_{n})$ &  $-\frac{i(-1)^n}{4a}(X_nX_{n+1}+Y_nY_{n+1})$ \\
    $\overline{\psi}\gamma_1\partial_1\psi$ & $-\frac{1}{2a^2}(\chi^\dagger_n\chi_{n+1}-\chi^\dagger_{n+1}\chi_{n})$ &  $-\frac{i}{4a^2}(X_nX_{n+1}+Y_nY_{n+1})$ \\
\end{tabular}
\end{center}
    \caption{The three distinct representations of the Dirac fermion field are formulated in a flat background. To incorporate the AdS black hole background, the redshift factor must be applied to the corresponding operators.}
    \label{tab:dic}
\end{table}

\subsection{Tensor-network simulations}
All results are obtained with open-boundary matrix-product-state (MPS) \cite{Perez-Garcia:2006nqo} simulations in Julia using \texttt{ITensors.jl} and \texttt{ITensorTDVP}. The lattice Hamiltonian is implemented as an \texttt{AutoMPO} and converted to an MPO. Ground states are computed by density matrix renormalization group (DMRG) \cite{RevModPhys.77.259} with 15 sweeps, a truncation cutoff $10^{-9}$, and bond dimension increased up to $\chi_{\rm gs}\le 1200$. Real-time evolution is performed with time-dependent variational principle (TDVP) \cite{PhysRevLett.107.070601} using $\Delta t=10^{-4}$ (up to $t_{\max}=5\times 10^{-2}$), with truncation cutoff $10^{-9}$ and a maximum bond dimension $\chi_{\max}=400$. Numerical stability was checked by halving $\Delta t$ and increasing $\chi_{\max}$, confirming no visible changes in the reported charge/current/entropy data.

\section{\label{sec:dynamics}Chiral Dynamics in Curved Spacetime}
In this section, we analyze the real-time dynamics of Dirac fermions in AdS$_2$ and AdS$_2$ black hole backgrounds, using the lattice Hamiltonian derived above. We show how curvature and the spin connection generate geometry-induced chiral transport, leading to asymmetric propagation of charge and entanglement. After introducing the quench protocol, we study charge separation, entanglement growth, and their confinement within an inhomogeneous LR cone. We also construct curvature-weighted correlation functions, to characterize the causal structure and chiral propagation of excitations.

\subsection{\label{chiral}Geometry Induced Chiral Waves}

The chiral nature of the dynamics arises from the AdS geometry itself, which induces a left-right imbalance in the propagation of fermionic modes. Through the spin connection, the curved background acts as a position-dependent \emph{chiral chemical potential} 
\begin{equation}
    \mu_5(z) = \frac{1}{2z},
\end{equation}
producing a geometrically sourced chiral asymmetry that persists even in the absence of external fields or interactions. This imbalance, encoded in the term \(\tfrac{1}{2z}\bar{\psi}\gamma^{1}\psi\) in the Hamiltonian \eqref{eq:Ham_AdS}, originates purely from the spacetime curvature and exists irrespective of the fermion mass. 

In \(1+1\) dimensions, the axial anomaly relation
\begin{equation}
    \partial_\mu J^\mu_5 = 2im\,\bar{\psi}\gamma_5\psi
\end{equation}
implies that the vector current \(J \equiv \bar{\psi}\gamma^{1}\psi\) is not conserved when \(m \neq 0\). In the AdS background this non-conservation is reinforced by the curvature induced \(\mu_5\), leading to asymmetric transport of charge and energy between left- and right-moving components. 

On the lattice, the same geometric coupling appears as an additional Dzyaloshinskii-Moriya type term \cite{Dzyaloshinskii1958,Moriya1960}
\begin{equation}
    D_n \left( X_n Y_{n+1} - Y_n X_{n+1} \right), \qquad D_n \propto \frac{n}{L^2},
\end{equation}
supplementing the redshift-weighted kinetic term
\begin{equation}
    J_n \left( X_n X_{n+1} + Y_n Y_{n+1} \right), \qquad J_n \propto \alpha_n^2.
\end{equation}
Together, these terms produce different local propagation speeds on the two sides of an excitation, giving rise to the asymmetric chiral gravitational waves observed in the simulations. In real-time evolution this manifests as left-right asymmetric charge and entanglement fronts, whose slopes depend on the local redshift factor \(\alpha(r)\), and are further modified by the spin connection driven chiral bias \(\mu_5(z)\).

\subsection{Chiral Entanglement Dynamics and Charge Separation}
\label{subsec:chiral-ent-and-charge}

To probe the geometry–induced chirality in real-time, we start from the ground state $\ket{\psi_{\mathrm{gs}}}$ of the curved space lattice Hamiltonian (\ref{eq:Ham_AdS}), and create a localized zero net charge dipole excitation:
\begin{equation}
\label{eq:dipole-state}
\ket{\psi_{\mathrm{di}}}\;:=\;X_{N/2}\,X_{N/2+1}\,\ket{\psi_{\mathrm{gs}}},
\end{equation}
where $X_i$ flips the spin at site $i$. Next, we will evolve $\ket{\psi_{\mathrm{di}}}$ in time, $\ket{\psi_{\mathrm{di}}(t)}=e^{-itH}\ket{\psi_{\mathrm{di}}}$. To isolate the signal generated by the dipole, all expectation values are background–subtracted:
\begin{equation}
\label{eq:subtract}
\Delta O(t) \equiv \langle{O(t)}\rangle_{\mathrm{di}}-\langle{O(t)}\rangle_{\mathrm{gs}} \ ,
\end{equation}
which cleanly reveals the geometry-induced chirality. We will study the expectation values of the charge profile (\ref{QC}), and the local entanglement entropy, obtained from MPS simulations.

We quantify entanglement by the bipartite von Neumann entropy across each MPS bond $b\in\{1,\ldots,N-1\}$ associated with the partition $\{1,\ldots,b\}\,|\,\{b\!+\!1,\ldots,N\}$. Writing the time-evolved pure state as $\ket{\psi(t)}$, we bring the MPS into canonical form with the orthogonality center at bond $b$, form the two-site tensor spanning the cut, and perform an SVD to obtain the Schmidt values $\{\lambda^{(b)}_k(t)\}_{k=1}^{\chi_b(t)}$, where $\chi_b(t)$ is the time‑dependent Schmidt rank (bond dimension) of the MPS across bond $b$ and time $t$. Equivalently, it is the number of nonzero Schmidt coefficients $\lambda^{(b)}_k(t)$. The local entanglement entropy then reads
\begin{equation}
\label{eq:Sb}
S_b(t)\;=\;-\sum_{k=1}^{\chi_b(t)}
\bigl[\lambda^{(b)}_k(t)\bigr]^2\,
\log\!\bigl([\lambda^{(b)}_k(t)]^2\bigr),
\qquad \sum_k [\lambda^{(b)}_k(t)]^2=1,
\end{equation}
and, because $\ket{\psi(t)}$ is pure, it is identical on the two sides of the cut.

While the bond entropy $S_b(t)$ across all cuts provides a space--time resolved entanglement profile naturally suited to MPS simulations, it is also useful to consider entanglement of small contiguous subsystems. For an interval $A=[i,i+\ell-1]$ we define the block von Neumann entropy
\begin{equation}
S_\ell(i,t)\equiv -\mathrm{Tr}\,\rho_A(t)\log\rho_A(t),\qquad
\rho_A(t)=\mathrm{Tr}_{\bar A}\,|\psi(t)\rangle\langle\psi(t)|,
\end{equation}
with $\ell=1,2,\dots$. In particular, the two-site entropy $S_2(i,t)$ tracks the build-up of short-range entanglement carried by the propagating fronts, complementing the bond entropy which quantifies entanglement across a cut. In the following we focus on $S_b(t)$ because it directly enters the Lieb--Robinson first-passage analysis and is the most directly accessible quantity in canonical MPS form, while $S_\ell(i,t)$ serves as a cross-check and a more local diagnostic.

\begin{figure}[H]
    \centering
    \includegraphics[width=0.49\linewidth]{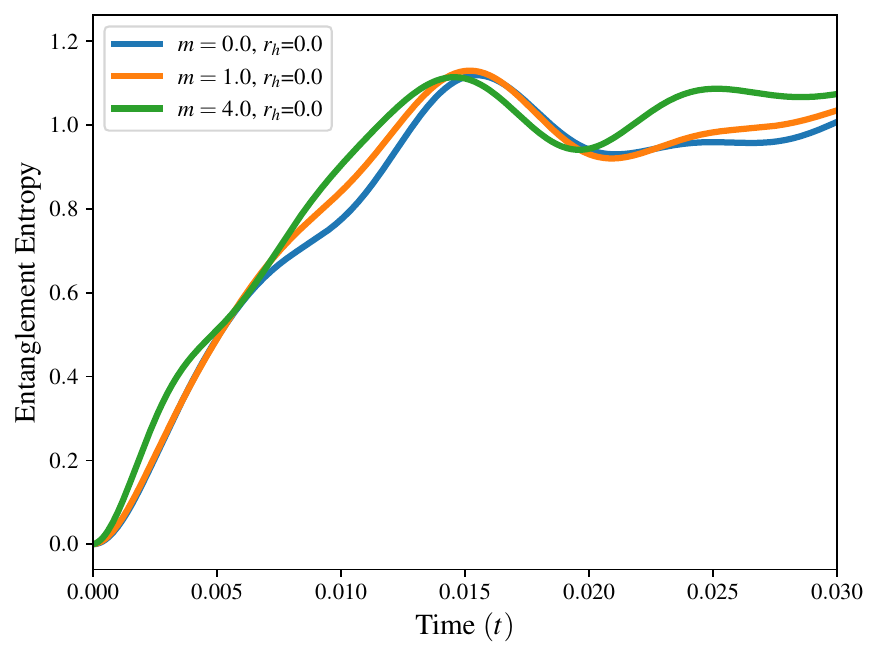}
    \includegraphics[width=0.49\linewidth]{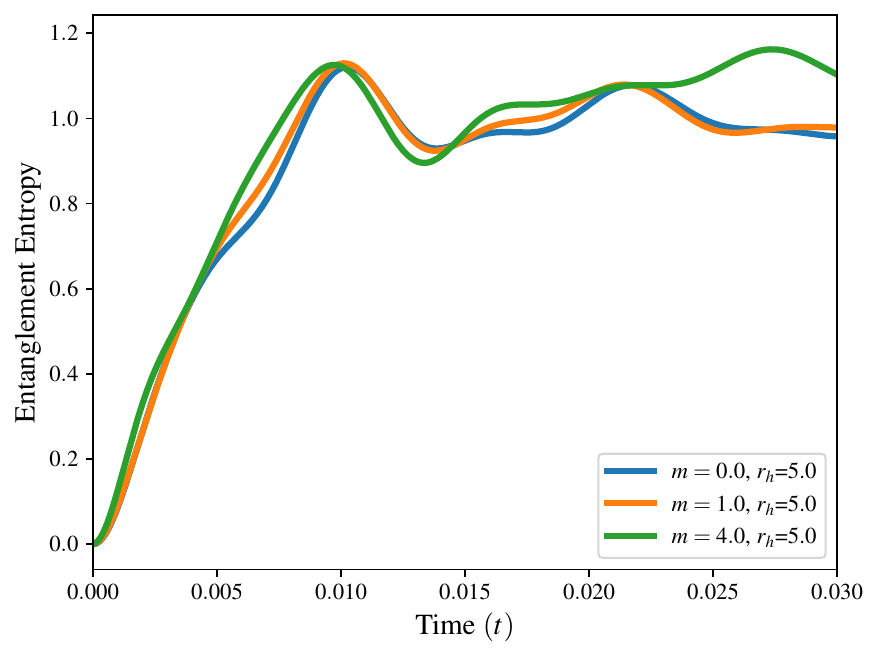}
    \caption{
Time evolution of the bipartite entanglement entropy $\Delta S_{L|R}(t)$ following a single-dipole quench in AdS$_2$ and AdS$_2$ black hole backgrounds ($N = 40$). Left: pure AdS ($r_h = 0$); right: AdS black hole ($r_h = 5$). Curves correspond to fermion masses $m = 0, 1, 4$. The initial linear rise of $\Delta S_{L|R}(t)$ reflects ballistic quasiparticle spreading confined within the LR cone, while the saturation is due to screening/dephasing in the finite inhomogeneous chain.}
\label{fig:EE}
\end{figure}

Figures~\ref{fig:dipole_charge_and_EE}–\ref{fig:EE} display the real-time charge profile 
$|{\Delta Q}_b(t)|$, and the corresponding local entanglement entropy $\Delta S_b(t)$, obtained from matrix product state (MPS) simulations at system size $N=40$.

Fig.~\ref{fig:dipole_charge_and_EE} shows the real-time evolution of the local charge (top) and the local entanglement entropy (bottom) for $m=0,1,4$ with $r_h=0$ and $N=40$. A similar simulation with the AdS black hole background is shown in Fig.~\ref{fig:dipole_charge_and_EE_BH}. The dynamics is manifestly left–right asymmetric, reflecting the position-dependent redshift in AdS that modifies local propagation speeds. The entanglement remains negligible outside the causal cone within numerical precision, consistent with quasiparticle spreading constrained by a LR velocity. Inside the cone, the growth rate and spread depend on the mass, with heavier $m$ exhibiting slower light-cone opening and reduced entanglement production. Introducing a horizon $r_h>0$, enhances the redshift, further suppressing front velocities and entanglement amplitudes relative to pure AdS.

\begin{figure}
    \centering
    \includegraphics[width=0.32\linewidth]{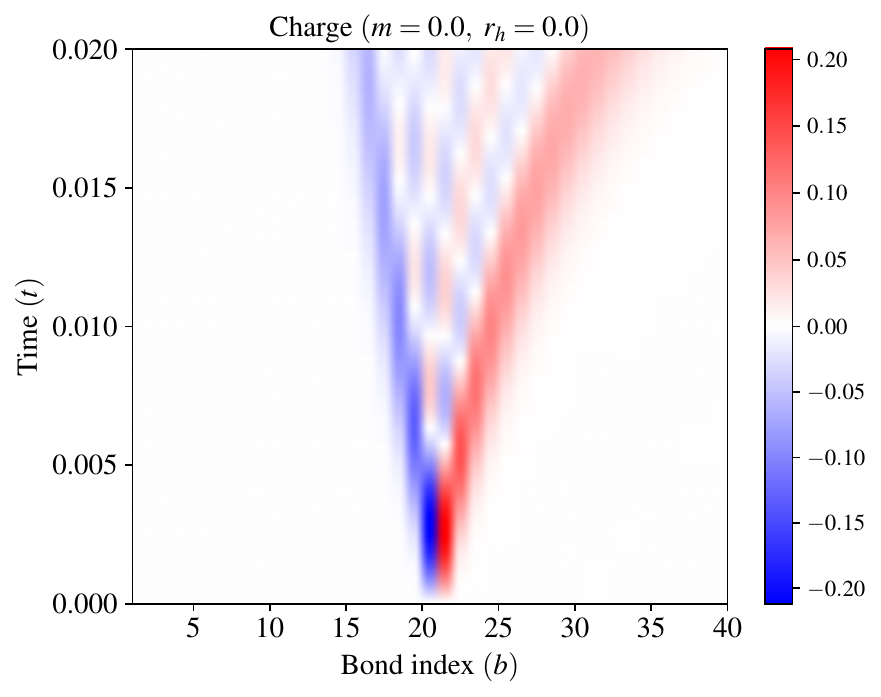}
    \includegraphics[width=0.32\linewidth]{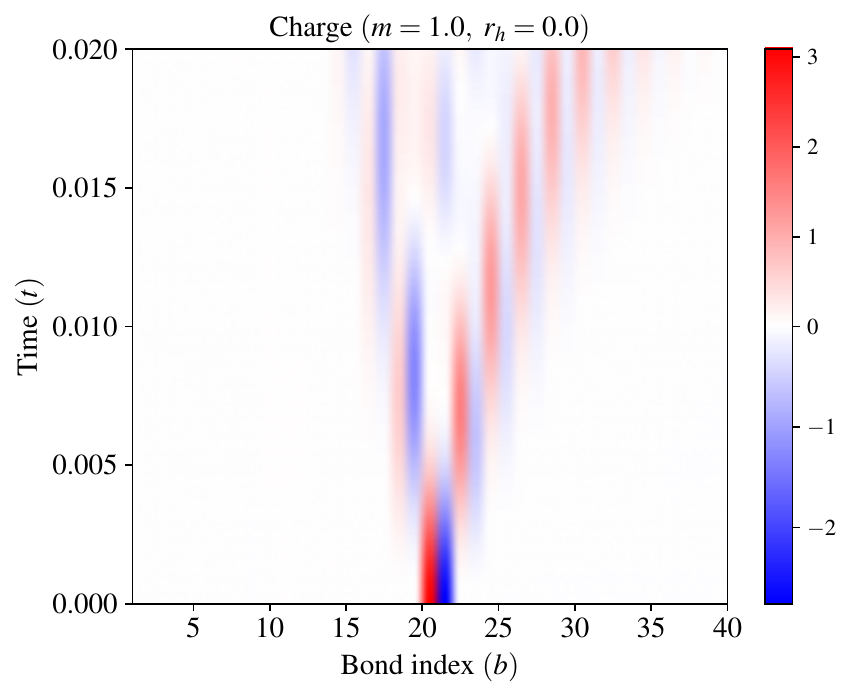}
    \includegraphics[width=0.32\linewidth]{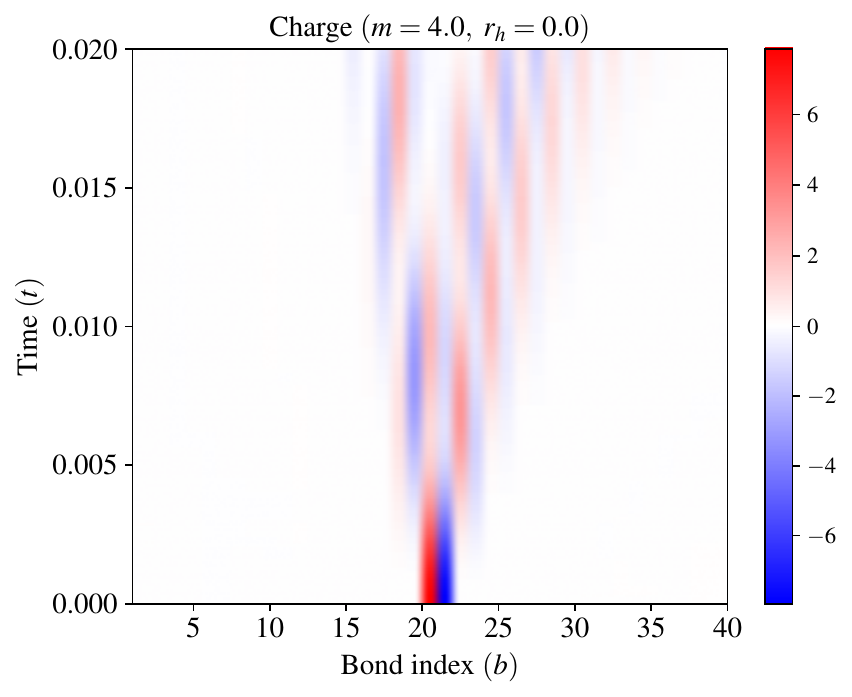}
    \includegraphics[width=0.32\linewidth]{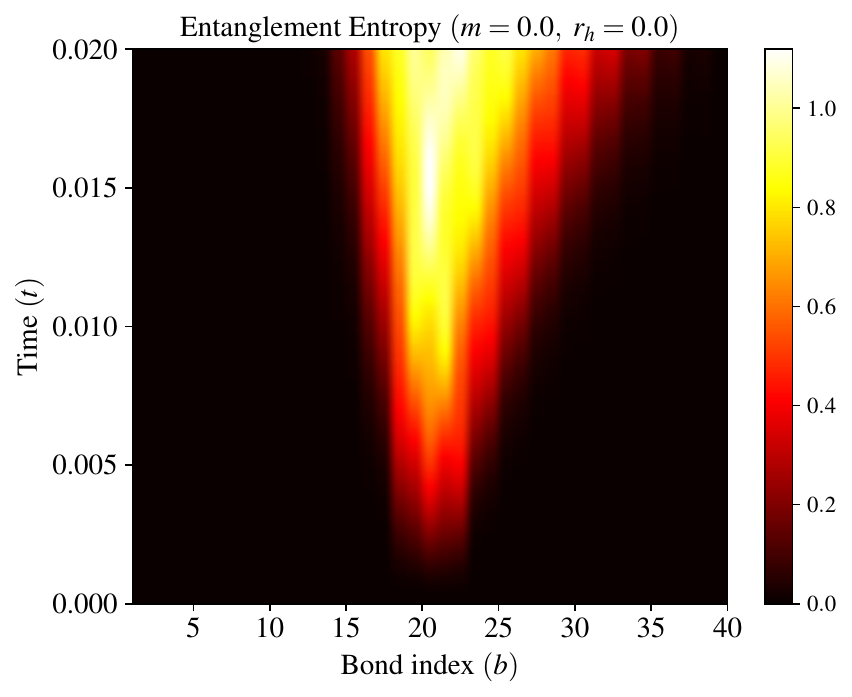}
    \includegraphics[width=0.32\linewidth]{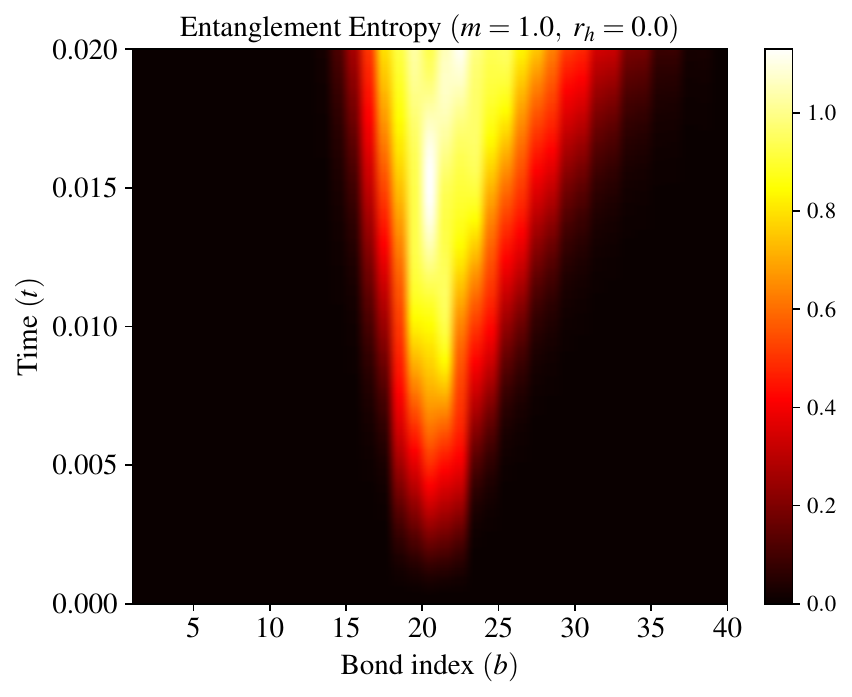}
    \includegraphics[width=0.32\linewidth]{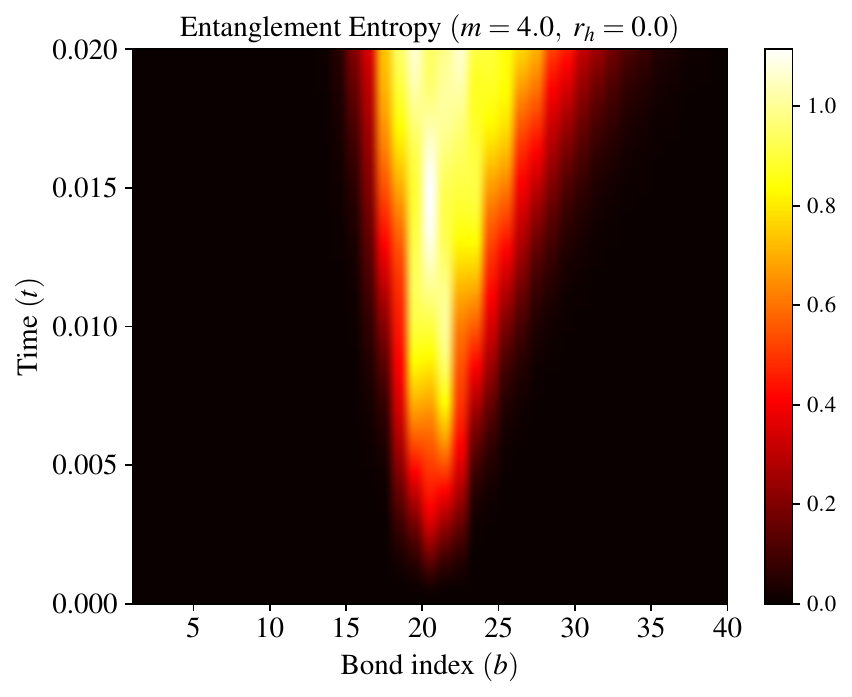}
    \caption{Real-time evolution of a single dipole excitation in pure AdS$_2$ showing chiral gravitational wave dynamics. Top panels display the local charge density, and bottom panels the corresponding local entanglement entropy, both simulated using MPS for a chain of $N = 40$ sites with $r_h = 0$. From left to right the fermion mass is $m = 0, 1, 4$. The outward tilted, left-right asymmetric wavefronts illustrate geometry induced chiral propagation arising from the AdS redshift and spin connection term, while entanglement growth remains confined within the inhomogeneous LR cone and slows with increasing $m$.  
    }
   \label{fig:dipole_charge_and_EE}
\end{figure}

\begin{figure}
    \centering
    \includegraphics[width=0.32\linewidth]{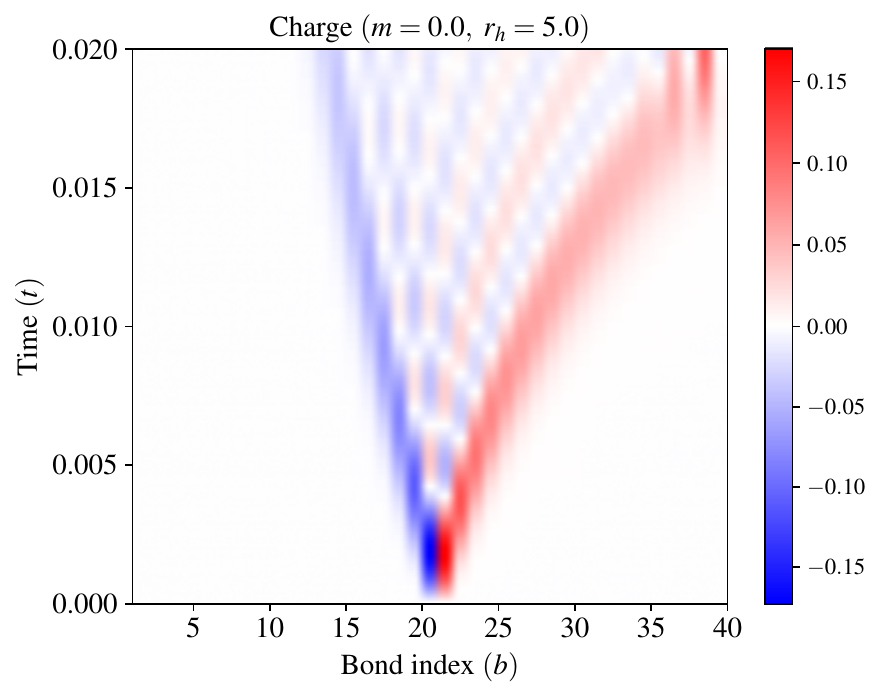
}
\includegraphics[width=0.32\linewidth]{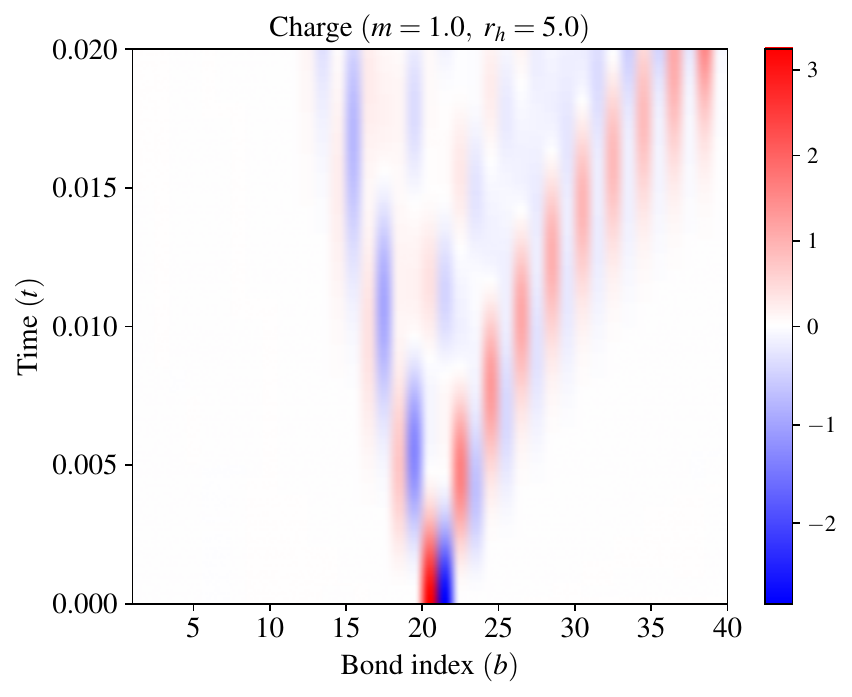
}
    \includegraphics[width=0.32\linewidth]{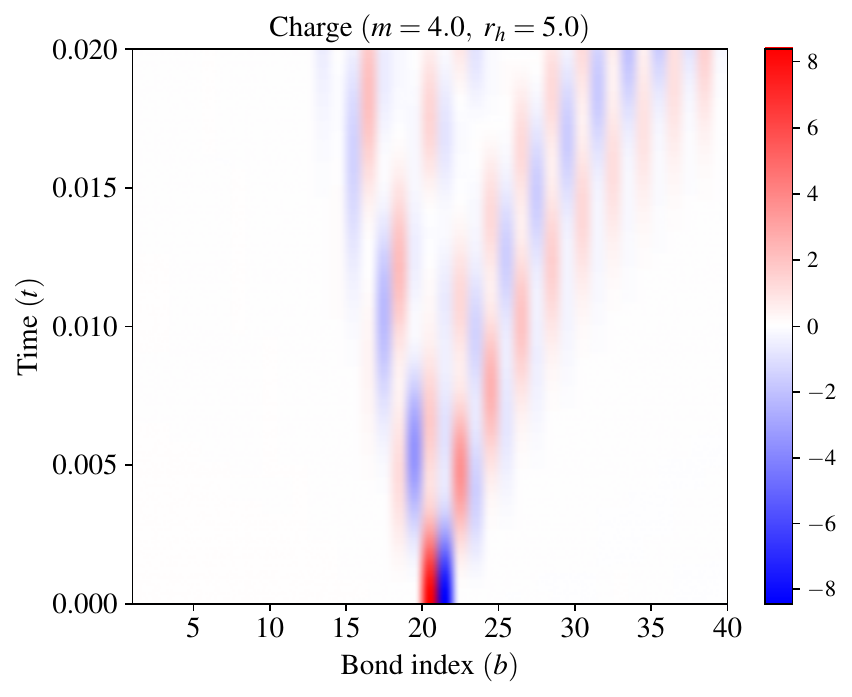
}
    \centering
    \includegraphics[width=0.32\linewidth]{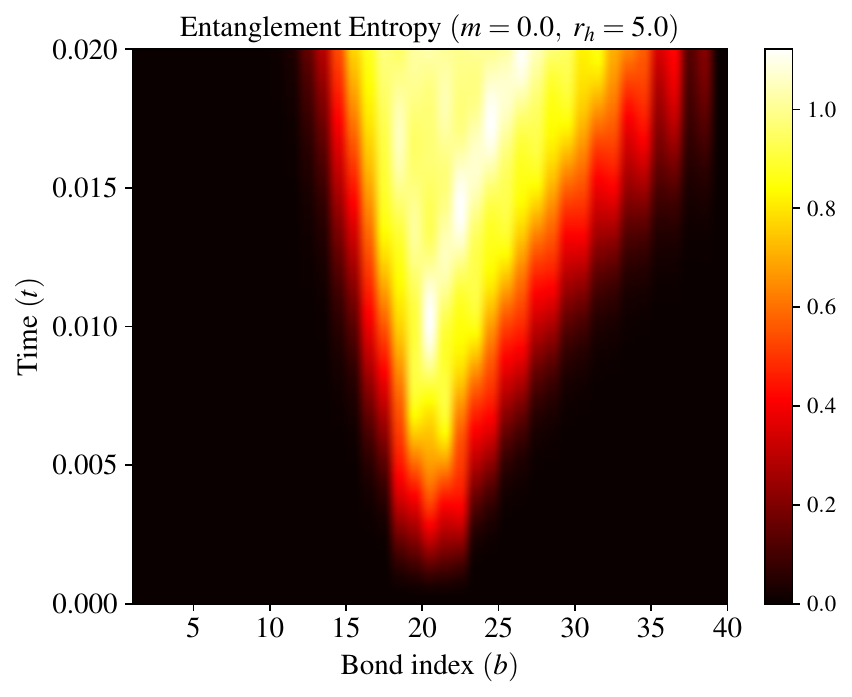
}
\includegraphics[width=0.32\linewidth]{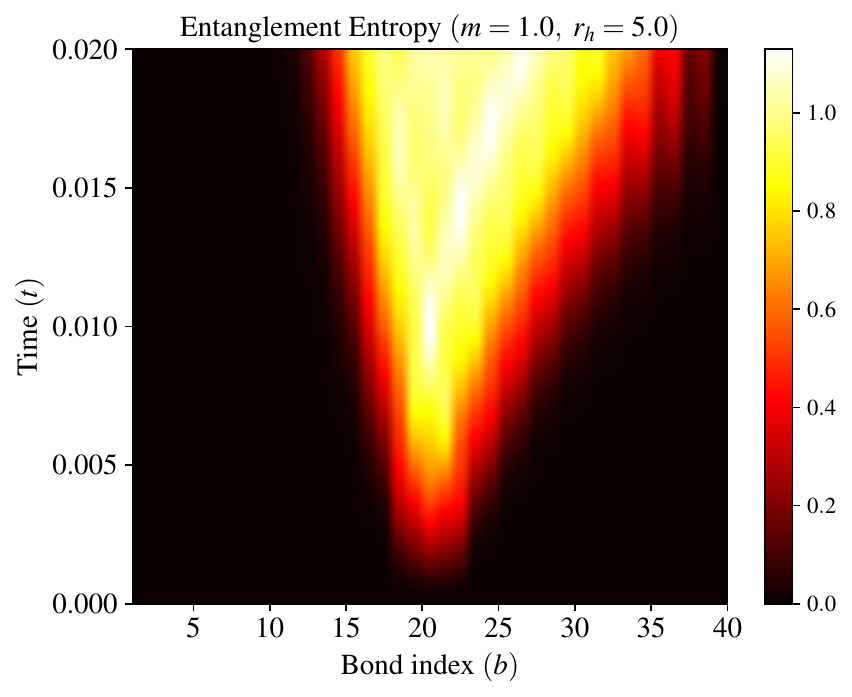
}
    \includegraphics[width=0.32\linewidth]{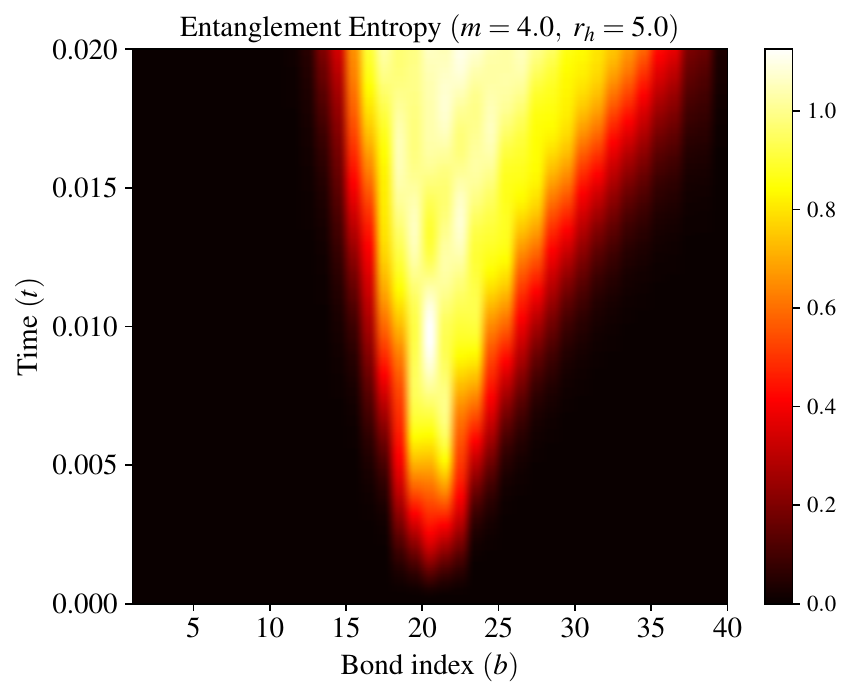
}
    \caption{
        Charge and entanglement entropy at horizon radius $r_h = 5$, with other parameters as in Fig.~\ref{fig:dipole_charge_and_EE}.  The presence of the horizon enhances the geometric redshift, producing slower, more asymmetric wavefronts and reduced entanglement growth compared to pure AdS. The smaller $b$ is, the closer it is to the event horizon.  %All signals remain confined within the inhomogeneous Lieb-Robinson cone, confirming causality-respecting chiral propagation.
    }
    \label{fig:dipole_charge_and_EE_BH}
\end{figure}

Fig.~\ref{fig:EE} shows the time evolution of the bipartite entanglement entropy across the central cut $\Delta S_{L|R}(t)$, following a local dipole quench in AdS$_2$. The entanglement entropy exhibits an initial growth, confined to a LR cone, and subsequently approaches a plateau. The central-cut entropy $\Delta S_{L|R}(t)$ exhibits an initial rise followed by a plateau. This late-time saturation is a generic feature of local quenches in finite one-dimensional systems and can be understood as a consequence of quasiparticle entanglement production followed by dephasing and screening in the inhomogeneous chain. In the present setup, one may view the dipole as creating counter-propagating particle--hole components whose separation initially increases the entanglement across the central cut, while at later times local excitations and interference reduce the rate of further long-range entanglement growth, leading to a saturation regime.

Note that for intuition we previously referred to this behavior as ``string breaking'', in analogy to (1+1)d gauge theories where separating charges nucleate screening pairs \cite{Florio:2024aix,PhysRevB.102.014308,De:2024smi}. We emphasize, however, that our model does not include a dynamical gauge field and therefore does not realize confinement or a literal flux tube. Accordingly, we interpret the observed plateau as screening-/dephasing-induced entanglement saturation and do not use it by itself as a standalone thermalization diagnostic.

\if{
To simulate the chiral dynamics in the AdS black hole background, we consider the real-time evolution of a single-dipole excitation,
\begin{equation}
\ket{\psi_{\mathrm{di}}}\coloneqq X_{N/2}\,X_{N/2+1}\ket{\psi_{\mathrm{gs}}}\,,
\end{equation}
where $X_i$ is a Pauli $X$ operator at site $i$, which flips the local spin there, and $\ket{\psi_{\mathrm{gs}}}$ denotes the ground state of the Hamiltonian \eqref{eq:Ham_AdS}. To isolate the dipole signal in the observables (e.g.\ charge or entanglement entropy), throughout this work, we remove the ground-state background by taking
\begin{equation}
\label{eq:subtract}
\Delta\langle{O(t)}\rangle \equiv \langle{O(t)}\rangle_{\mathrm{di}}-\langle{O(t)}\rangle_{\mathrm{gs}}.
\end{equation}

Fig.~\ref{fig:dipole_charge_and_EE} shows the real-time evolution of the local charge (top) and the local entanglement entropy (bottom) for $m=0,1,4$ with $r_h=0$ and $N=40$. A similar simulation with the AdS black hole background is shown in Fig.~\ref{fig:dipole_charge_and_EE_BH}. The dynamics is manifestly left–right asymmetric, reflecting the position-dependent redshift in AdS that modifies local propagation speeds. The entanglement remains negligible outside the causal cone within numerical precision, consistent with quasiparticle spreading constrained by a LR velocity. Inside the cone, the growth rate and spread depend on the mass, with heavier $m$ exhibiting slower light-cone opening and reduced entanglement production.

We consider the spreading of quantum correlations of the bipartite von Neumann entanglement entropy across every bond of the matrix
product state (MPS). For a chain of length $N$, each MPS bond $b\in\{1,\dots,N-1\}$ represents the bipartition
$\{1,\dots,b\}\; \big|\; \{b{+}1,\dots,N\}$. At (discrete) time $t_i$ the color value at row $b$ and column $i$ is the entropy:
\begin{equation}
S_b(t_i)\;=\; -\sum_{k=1}^{\chi_b(t_i)}
\Big(\lambda_k^{(b)}(t_i)\Big)^2\,
\log\!\Big(\lambda_k^{(b)}(t_i)\Big)^2,
\end{equation}
where $\{\lambda_k^{(b)}(t_i)\}_{k=1}^{\chi_b(t_i)}$ are the Schmidt coefficients of the pure state $\ket{\psi(t_i)}$ across the cut $b|b{+}1$, with $\sum_k \big(\lambda_k^{(b)}(t_i)\big)^2 = 1$, and $\chi_b(t_i)$ the Schmidt rank (bond dimension) at that cut. Because the global state is pure, the entropies of the two sides coincide, $S\!\left(\rho^{(b)}_{L}(t_i)\right) = S\!\left(\rho^{(b)}_{R}(t_i)\right)$. $S_b(t_i)$ is obtained by first bringing the MPS into canonical form with the orthogonality center at bond $b$, forming the two-site tensor across the cut, and performing an SVD to extract the singular values $\{\lambda_k^{(b)}(t_i)\}$. These singular values are the square roots of the nonzero eigenvalues of either reduced density matrix $\rho^{(b)}_{L}(t_i)=\Tr_R \ket{\psi(t_i)}\bra{\psi(t_i)}$ or $\rho^{(b)}_{R}(t_i)=\Tr_L \ket{\psi(t_i)}\bra{\psi(t_i)}$.

\begin{figure}[H]
    \centering
    \includegraphics[width=0.32\linewidth]{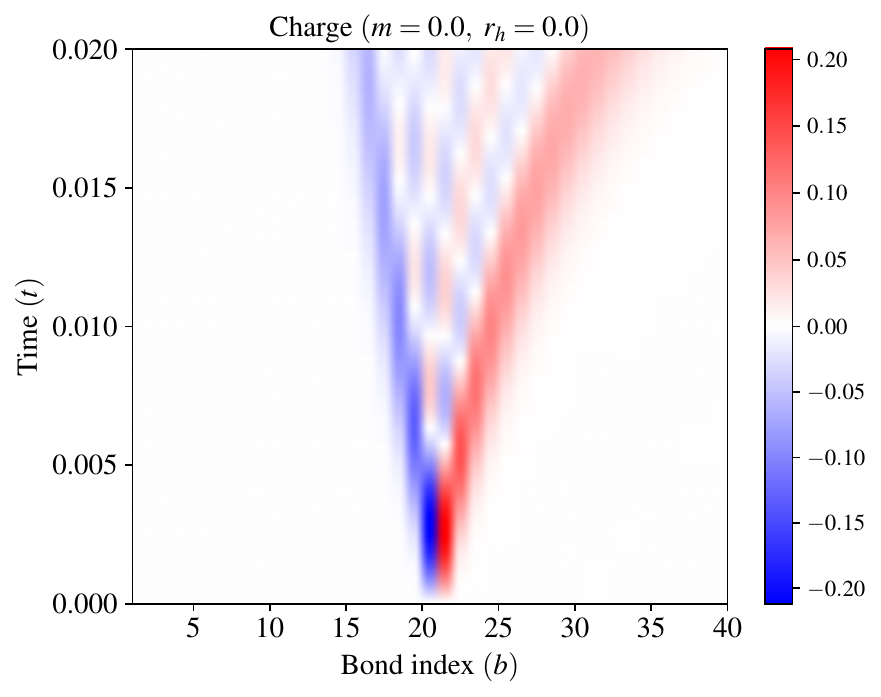}
    \includegraphics[width=0.32\linewidth]{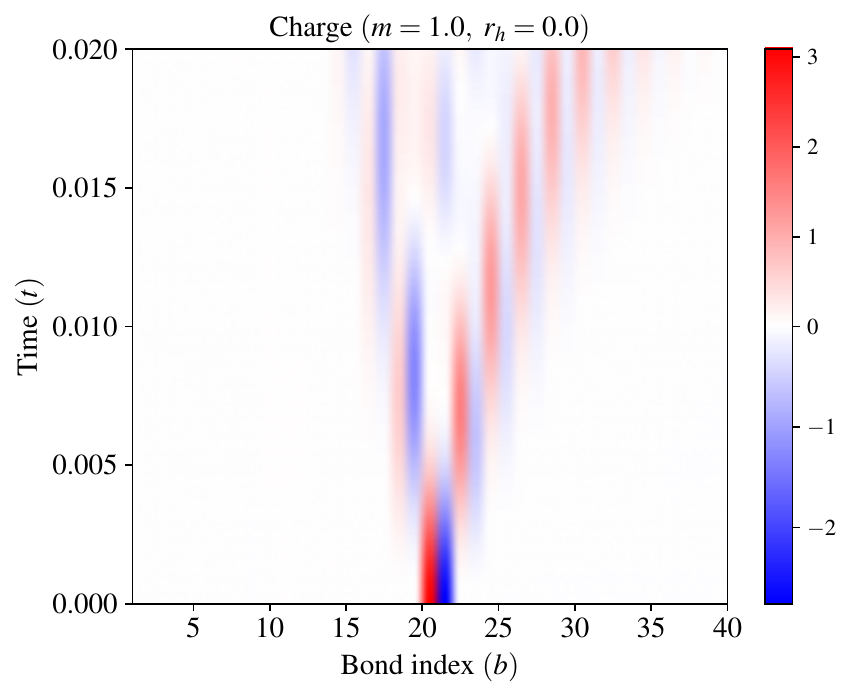}
    \includegraphics[width=0.32\linewidth]{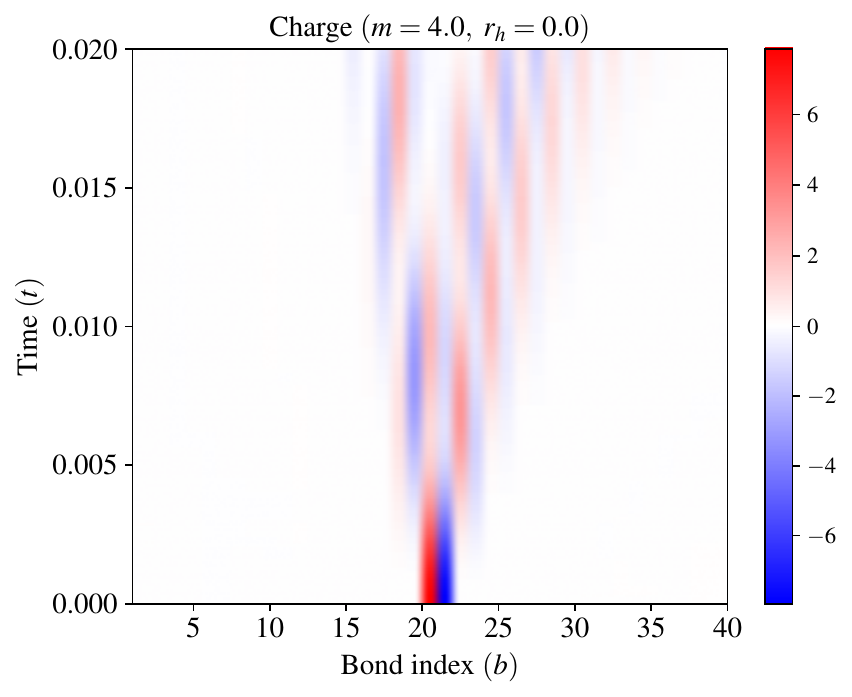}
    \includegraphics[width=0.32\linewidth]{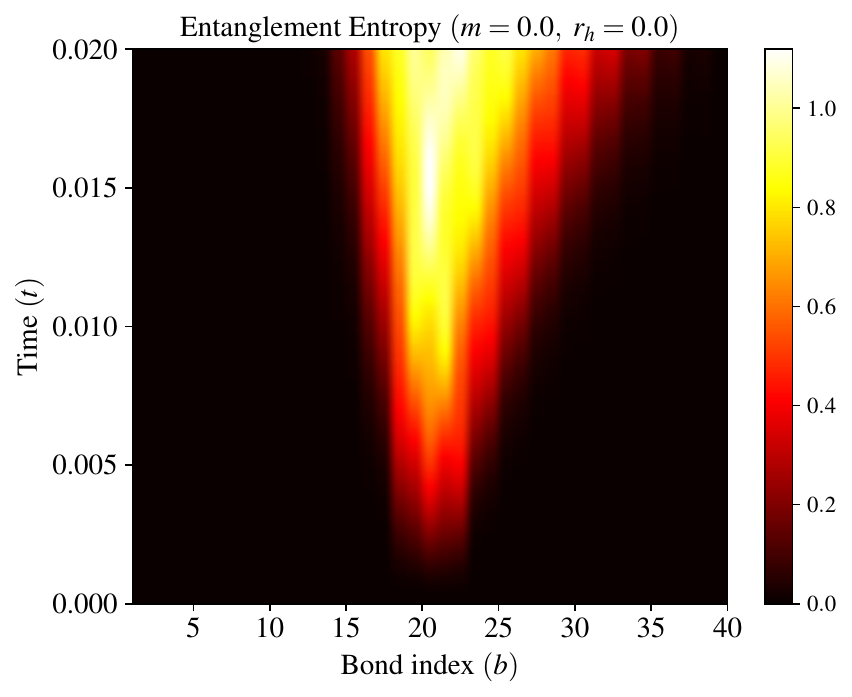}
    \includegraphics[width=0.32\linewidth]{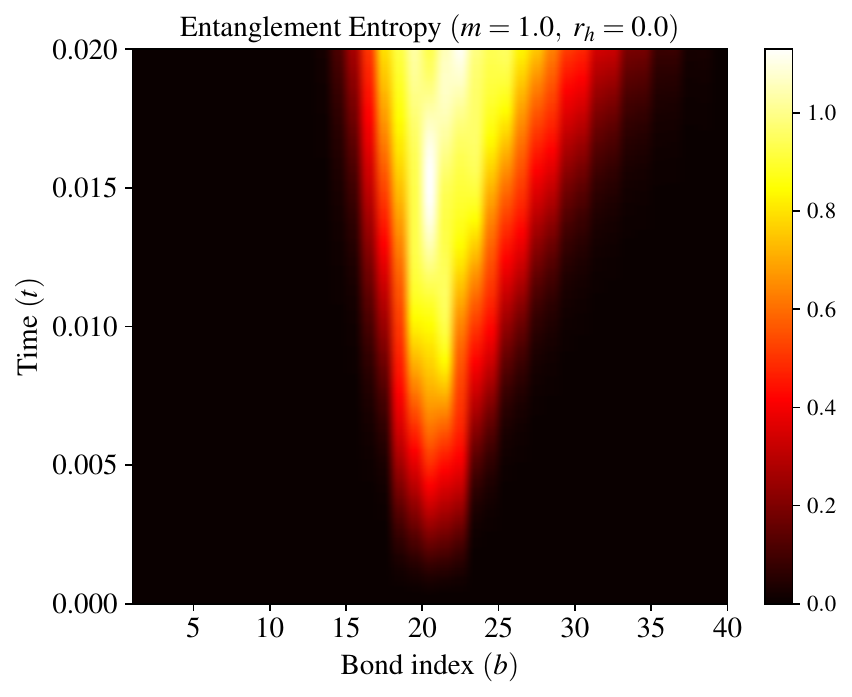}
    \includegraphics[width=0.32\linewidth]{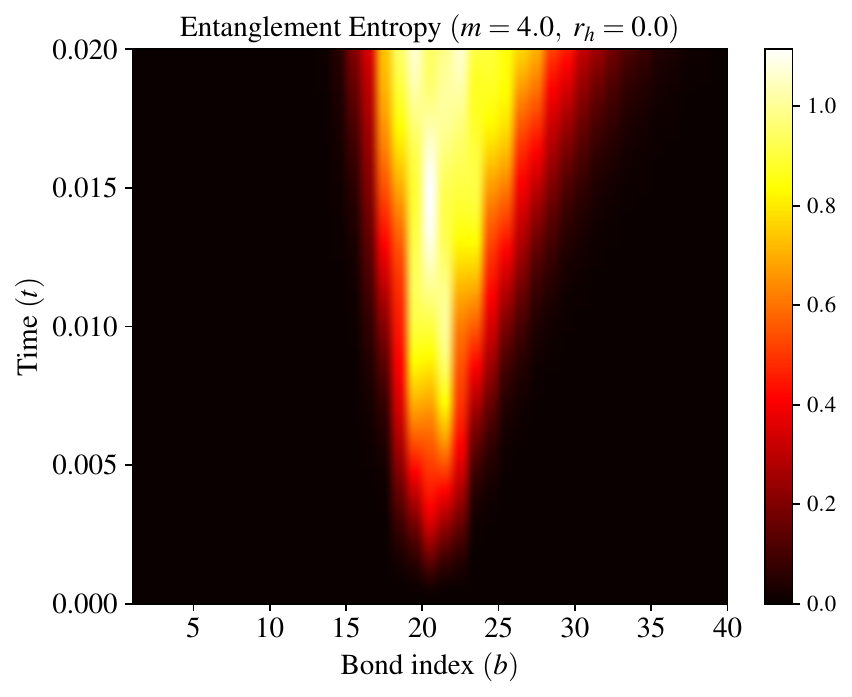}
    \caption{Real-time evolution of a single dipole excitation in pure AdS$_2$ showing chiral gravitational wave dynamics. Top panels display the local charge density, and bottom panels the corresponding local entanglement entropy, both simulated using MPS for a chain of $N = 40$ sites with $r_h = 0$. From left to right the fermion mass is $m = 0, 1, 4$. The outward tilted, left-right asymmetric wavefronts illustrate geometry induced chiral propagation arising from the AdS redshift and spin connection term, while entanglement growth remains confined within the inhomogeneous LR cone and slows with increasing $m$.
    }
   \label{fig:dipole_charge_and_EE}
\end{figure}

\begin{figure}[H]
    \centering
    \includegraphics[width=0.32\linewidth]{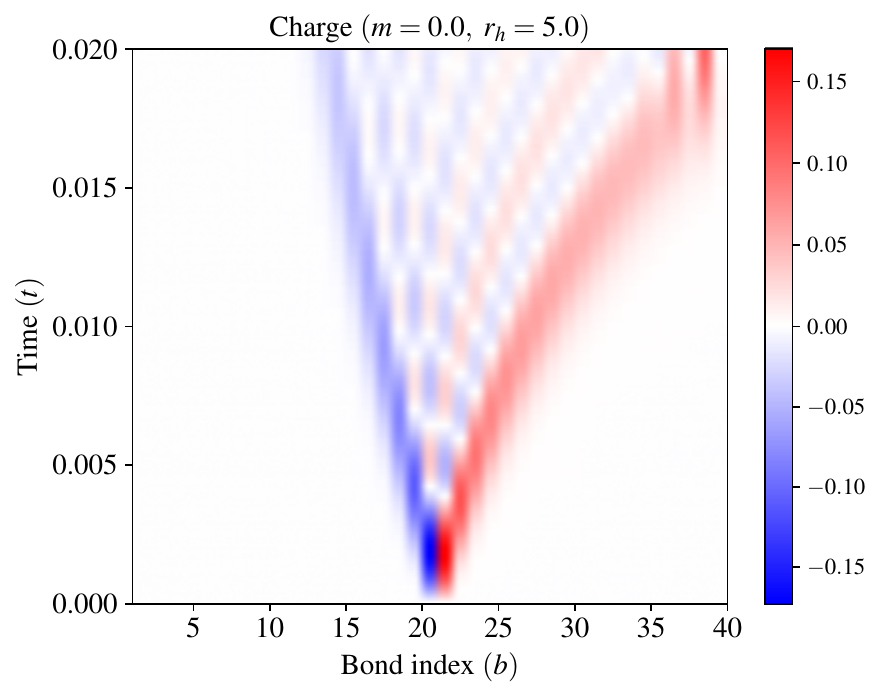}
\includegraphics[width=0.32\linewidth]{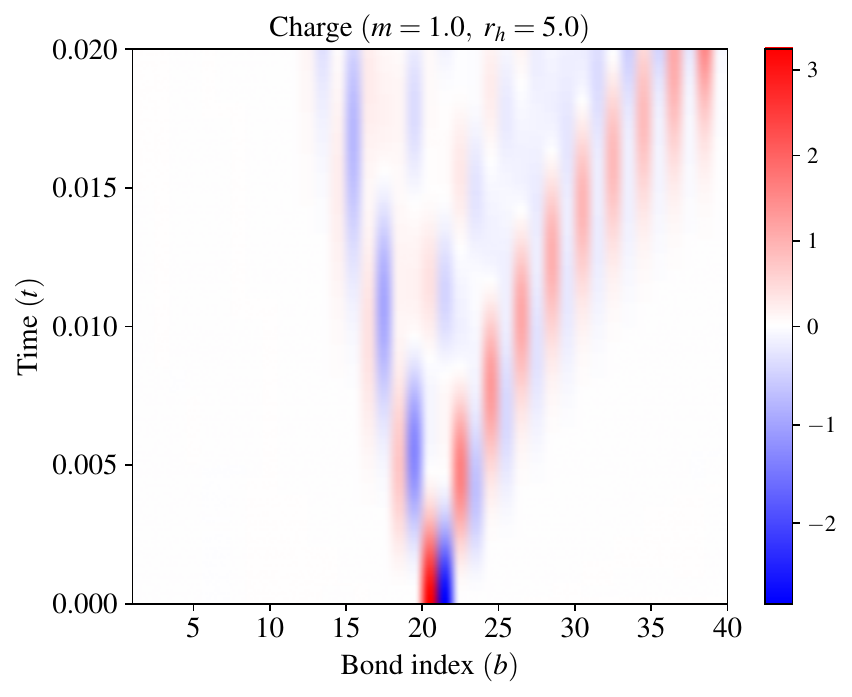}
    \includegraphics[width=0.32\linewidth]{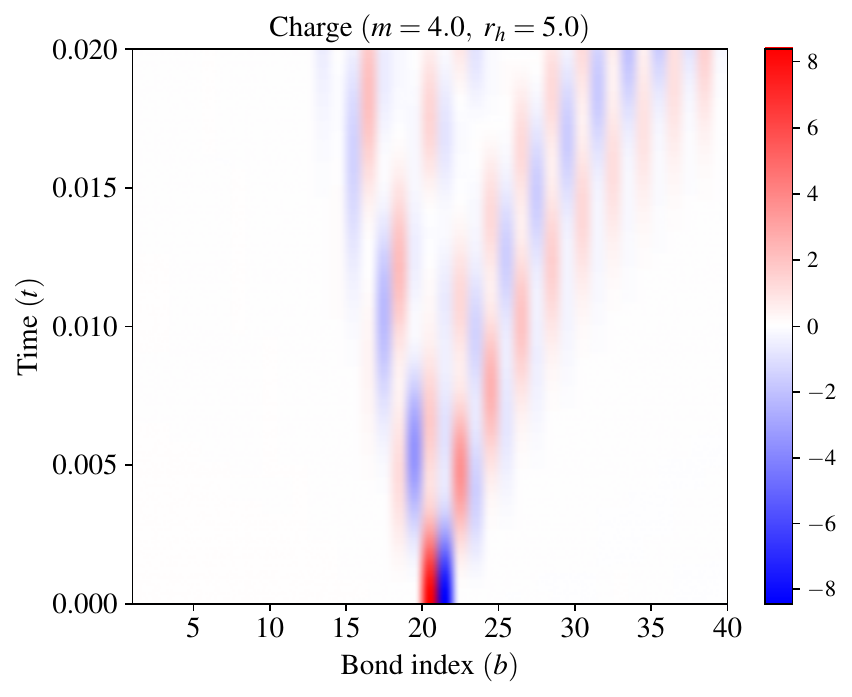}
    \centering
    \includegraphics[width=0.32\linewidth]{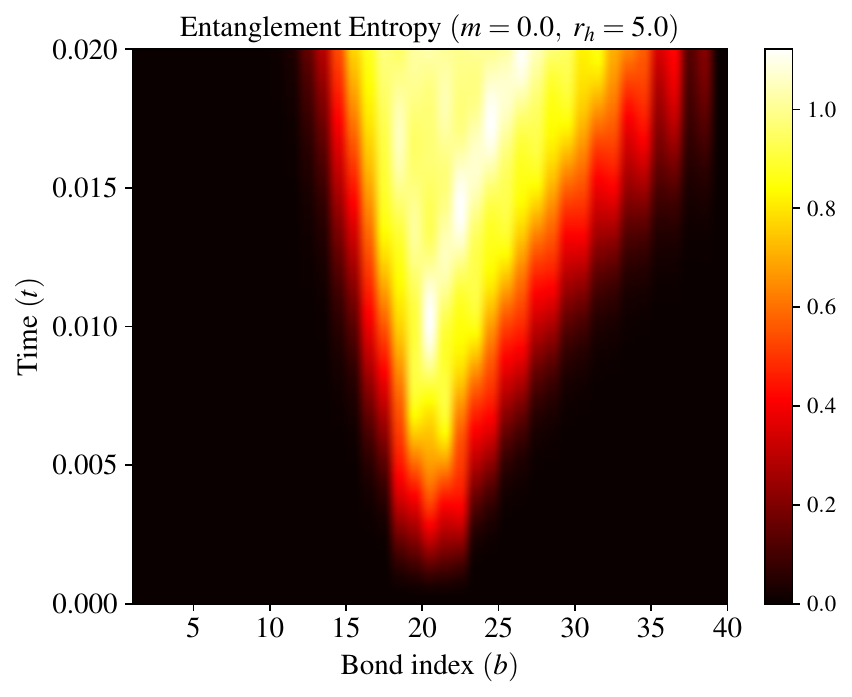}
\includegraphics[width=0.32\linewidth]{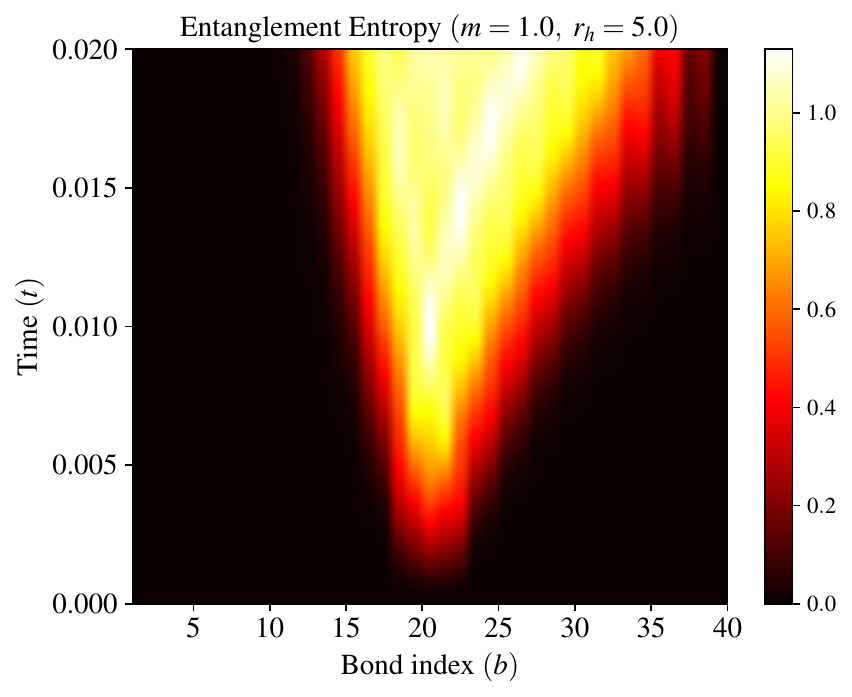}
    \includegraphics[width=0.32\linewidth]{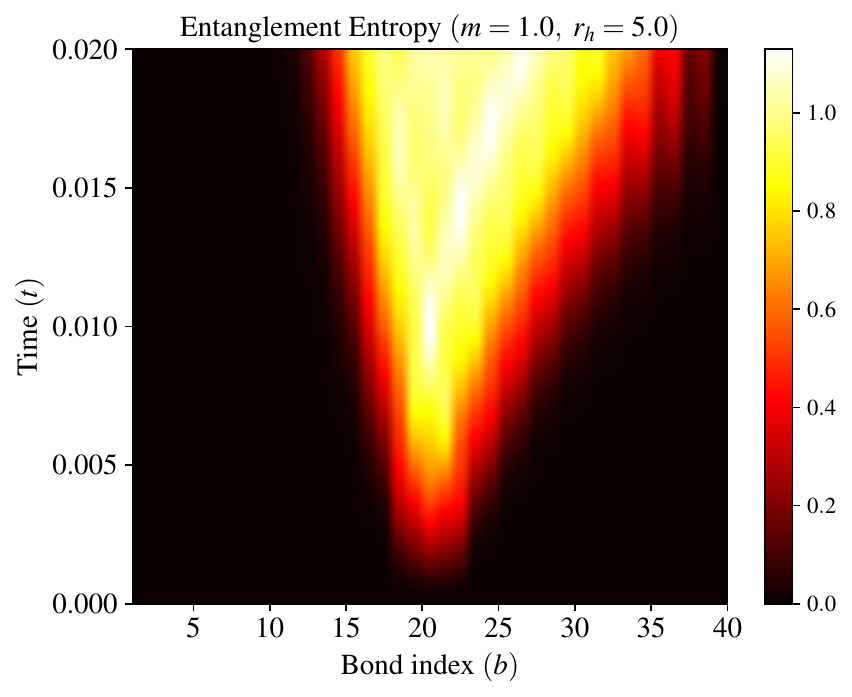}
    \caption{
        Charge and entanglement entropy at horizon radius $r_h = 5$, with other parameters as in Fig.~\ref{fig:dipole_charge_and_EE}.  The presence of the horizon enhances the geometric redshift, producing slower, more asymmetric wavefronts and reduced entanglement growth compared to pure AdS. %All signals remain confined within the inhomogeneous Lieb-Robinson cone, confirming causality-respecting chiral propagation.
    }
    \label{fig:dipole_charge_and_EE_BH}
\end{figure}

Fig.~\ref{fig:EE} shows the time evolution of the bipartite entanglement entropy $S_{L|R}(t)$ following a local dipole quench in AdS$_2$. The entropy exhibits an initial growth, confined to a LR cone (as shown in Figs.~\ref{fig:LB}), and subsequently approaches a plateau. This behavior admits a natural ``string‑breaking'' interpretation for a localized fermion–antifermion pair within the dipole: (i) at early times the pair separates ballistically and a domain‑wall–like string forms between them, accumulating energy; (ii) beyond a characteristic separation, the system lowers its energy by nucleating additional local excitations that screen the charges and effectively break the string. The resulting screening curtails the production of long‑range entanglement and drives the entropy toward saturation, consistent with the onset of local thermalization. 
}\fi

Because the AdS redshift enters the lattice description as spatially varying couplings, a bent (inhomogeneous) causal fan is an expected consequence of the Hamiltonian and serves as a useful consistency check. The nontrivial aspects we focus on are (i) the
chiral left--right asymmetry induced by the spin-connection term (an effective $\mu_5$) even in the absence of gauge fields, (ii) the quantitative dependence of front velocities on the fermion mass $m$ and horizon scale $r_h$, and (iii) the fact that both charge and entanglement growth are quantitatively confined by an inhomogeneous Lieb--Robinson arrival-time bound constructed from the local couplings. For comparison, in homogeneous flat-space chains the Lieb--Robinson cone is straight and parity-symmetric, and generating chirality typically requires external fields/chemical potentials or interactions \cite{AnthonyChen:2023bbe}.

These properties become more evident by comparing the dynamics in AdS and flat spaces. Chiral dynamics in a variety of $(1+1)$-dimensional Dirac systems in flat space have been extensively explored, with the Schwinger model (QED$_2$) serving as a canonical benchmark for gauge theories and as a particularly relevant target for near-term quantum devices \cite{PhysRev.128.2425,Klco:2018kyo,PhysRevD.109.114510,PRXQuantum.5.020315}. As explained above, the spin connection in the AdS Hamiltonian generates an effective chiral chemical potential $\mu_5$ even in the absence of a gauge field. This mechanism plays a role analogous to that of a magnetic field and therefore suggests notable parallels with flat space setups.\footnote{Strictly speaking, in QED$_2$ the magnetic field can be removed by gauge choice, so many analyses focus solely on the electric field. Flat space studies that introduce an explicit $\mu_5$ include \cite{PhysRevResearch.2.023342,Czajka2022,Ikeda:2024rzv}.} In \cite{PhysRevD.108.074001}, in contrast to the asymmetric fronts in Figs.~\ref{fig:dipole_charge_and_EE} and \ref{fig:dipole_charge_and_EE_BH}, \emph{symmetric} propagation of charge and current was reported for chiral magnetic waves \cite{Kharzeev:2010gd}. Relatedly, string breaking in QED$_2$ driven by an external time dependent source mimicking quark jets, was analyzed in \cite{2023arXiv230111991F,Florio:2024aix}, where plots analogous to Fig.~\ref{fig:EE} were used to discuss thermalization and hadronization.

\subsection{\label{sec:LR}Lieb–Robinson Bound}

To quantify the causal spreading of correlations, we use a first-passage analysis of the bond entanglement. For a small entropy threshold $\varepsilon_S$ we define
\begin{equation}
\label{eq:t_E}
  t^{(E)}_b \coloneqq \min\{\,t:\; \Delta S_b(t)\ge\varepsilon_S \,\},
\end{equation}
and determine directional entanglement velocities by fitting $b$ versus $t^{(E)}_b$ separately to the left ($b<b_0$) and right ($b>b_0$) of the quench bond $b_0=N/2$, yielding $v_E^{\mathrm{L}}$ and $v_E^{\mathrm{R}}$. This procedure extracts the ballistic front associated with the earliest sustained rise of $\Delta S_b(t)$, and provides a model-independent measure of the left-right asymmetry. 

Write the two-site term on bond $(n,n+1)$ in our Hamiltonian as:
\begin{equation}
\label{eq:hopping_term}
    h_{n,n+1}=J_n(X_nX_{n+1}+Y_nY_{n+1}) + D_n(X_nY_{n+1}-Y_nX_{n+1})
\end{equation}
with $J_n=\alpha_n^2/(4a)$ and $D_n=a n/(8L^2)$. Following standard LR constructions \cite{Lieb:1972wy,PhysRevLett.99.167201,PhysRevLett.113.127202,PRXQuantum.4.020349,PhysRevLett.119.100601}, we define a local speed scale $v_{\rm loc}(n)=\kappa\,\|h_{n,n+1}\|$. In the plots we set $\kappa=1$, so $v_{\rm loc}(n)=\|h_{n,n+1}\|=2\sqrt{J_n^2+D_n^2}$.

Beyond serving as a consistency check, the inhomogeneous LR arrival-time bound provides a quantitative \emph{arrival-time predictor} built solely from the local couplings. We use it in two ways: (i) to define first-passage times for the entanglement front and extract directional velocities, and (ii) in the two-dipole protocol (Sec.~\ref{sec:scattering}) to predict when the inward causal branches reach the central cut and therefore when central bipartite entanglement can begin to grow.

\begin{figure}[H]
    \centering
    \includegraphics[width=0.49\linewidth]{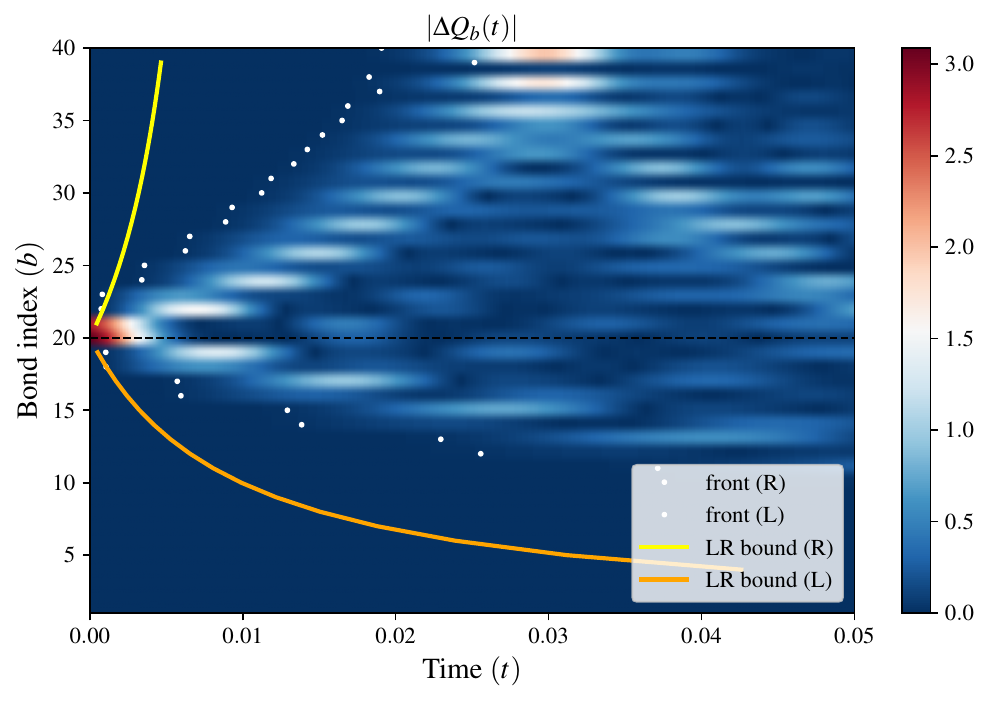}
    \includegraphics[width=0.49\linewidth]{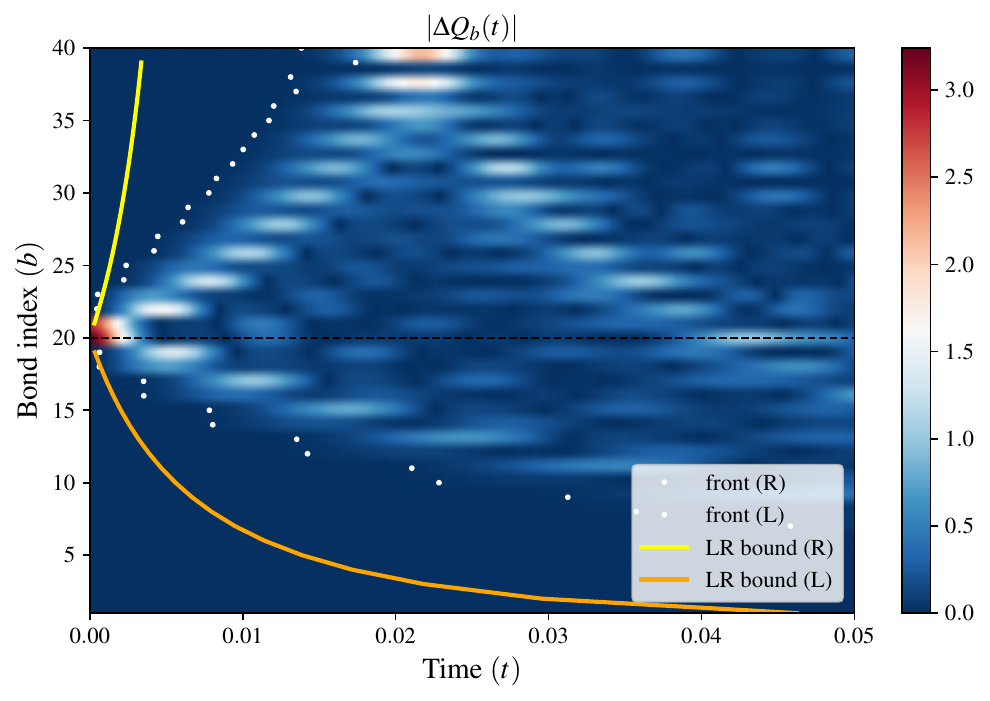}
    \includegraphics[width=0.49\linewidth]{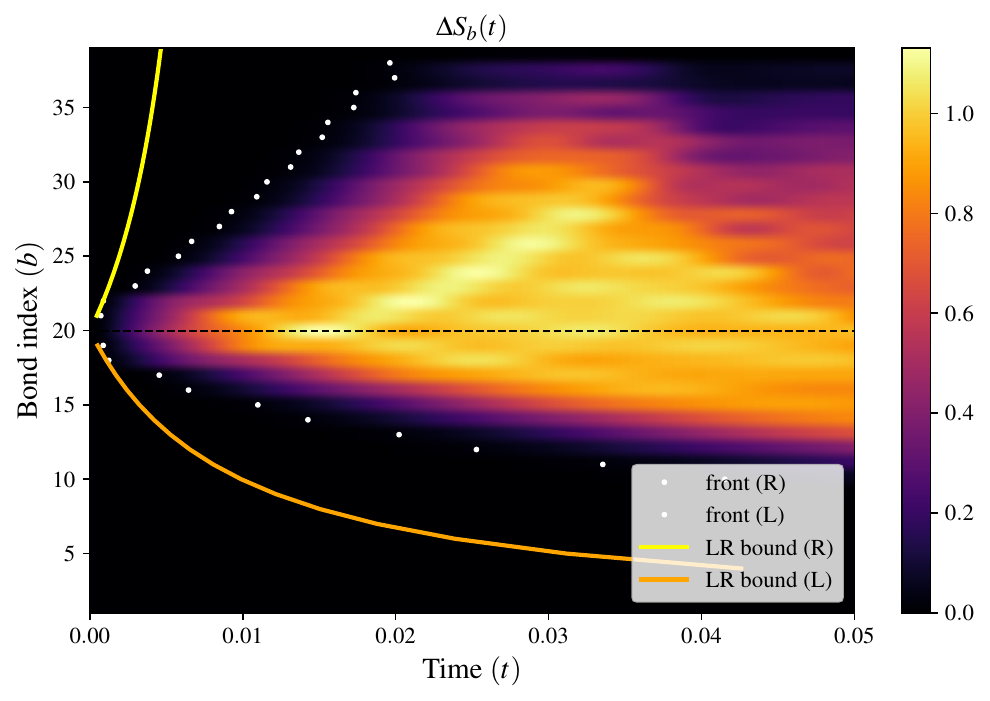}
    \includegraphics[width=0.49\linewidth]{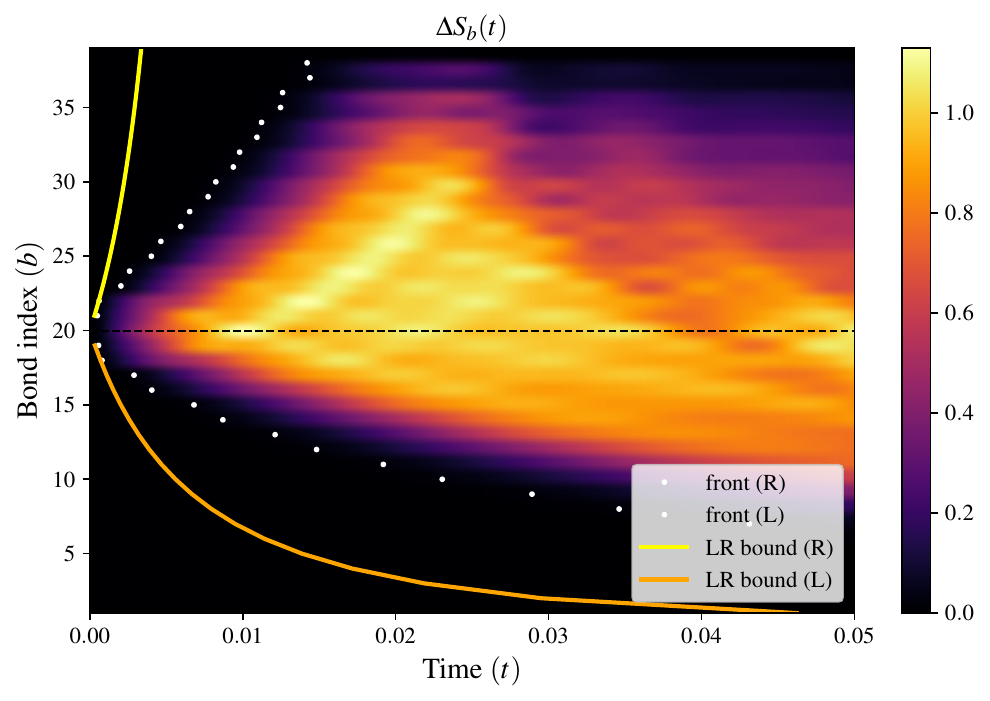}
    \caption{Heatmaps of the charge $|\Delta Q_b(t)|$ (top) and the entanglement entropy $\Delta S_b(t)$ (bottom) with first–passage markers (white dots) and inhomogeneous LR arrival–time bounds from the quench bond $b_0$ (yellow/orange for right/left). All measured fronts lie strictly inside the bound, with a clear chiral left–right asymmetry in the slopes. From left to right: $r_h=0,5$. ($N=40,m=1$) For $r_h \neq 0$, a smaller $b$ means it is closer to the event horizon, where the waves propagate slower.}
    \label{fig:LB}
\end{figure}

The resulting inhomogeneous arrival-time bound from the quench bond $b_0$ to bond $b$ is
\begin{equation}
t_{\rm bound}(b_0\!\to\!b) = \sum_{\ell\in {\rm path}} \frac{1}{v_{\rm loc}(\ell)} .
\end{equation}
We evaluate $t_{\mathrm{bound}}$ piecewise on the left/right of $b_0$, and overlay the curve $t=t_{\mathrm{bound}}$ on the heatmaps of $|{\Delta Q}_b(t)|$ and $\Delta S_b(t)$ (Fig.~\ref{fig:LB}). One can obtain heatmaps of the current showing similar properties. In all cases the first-passage markers lie strictly inside the LR cone, with clear and quantitative left--right asymmetry in the slopes; increasing $r_h$ reduces front speeds near the event horizon compared to those near the boundary, consistent with the redshifted couplings.

In the dipole-dipole setting (Sec.~\ref{sec:scattering}), the same bound constructed from $v_{\mathrm{loc}}$ predicts the arrival of the inward LR branches at the central cut. The onset of the central entropy $\Delta S_{L|R}(t)$ occurs when these branches meet, consistent with the LR overlays on the local heatmaps (Fig.~\ref{fig:two_source_LR}).

\subsection{\label{sec:corr}Correlation Functions}
We consider the following correlation function \cite{Kubo:1957mj,Cheneau:2012zdz,Bernecker:2011gh,Son:2002sd,Barata:2024apg,Barata:2024bzk}:
\begin{equation}
    \Pi^{\mu\nu}(t;x_1,x_0)=\langle J^{\mu}(t,x_1)J^\nu(t_0,x_0)\rangle,\quad t_0=0,
\end{equation}
with $\mu=0,1$ denoting the charge and current operators, respectively. On the lattice we use the weighted charge and current operators \eqref{QC}.

The current operator is constructed from the same bond terms \eqref{QC} that control local propagation. For this reason $\Pi^{11}$ is typically more sensitive than $\Pi^{00}$ to changes in the local couplings induced by the redshift and spin-connection (DM) terms. The leading peak in $|\Pi^{\mu\nu}(t;x_1,x_0)|$ provides a time-of-flight marker for the front arrival at $x_1$, while the subsequent oscillations reflect post-front interference and dephasing/screening in the finite inhomogeneous chain.

The localized dipole acts as a source of counter-propagating quasiparticles whose velocities are modulated by the AdS redshift. The right-moving packet propagates from $x_0$ to $x_1$ and imprints a peak in the correlators when it arrives. The subsequent oscillations track post-front interference and screening. The same chiral propagation and causal confinement are seen in the charge/entropy heatmaps (Figs.~\ref{fig:dipole_charge_and_EE} and \ref{fig:dipole_charge_and_EE_BH}), and the correlator peaks provide a concise, complementary diagnostic of that wave.

\begin{figure}[H]
    \centering
    \includegraphics[width=\linewidth]{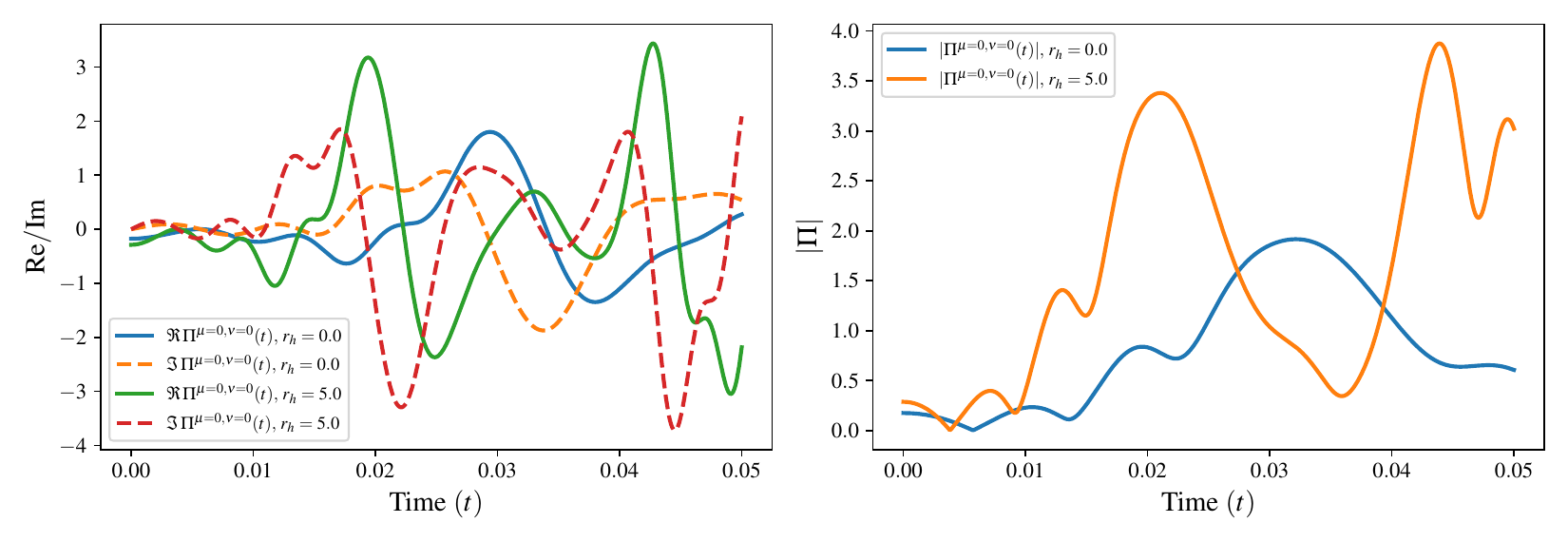}
    \includegraphics[width=\linewidth]{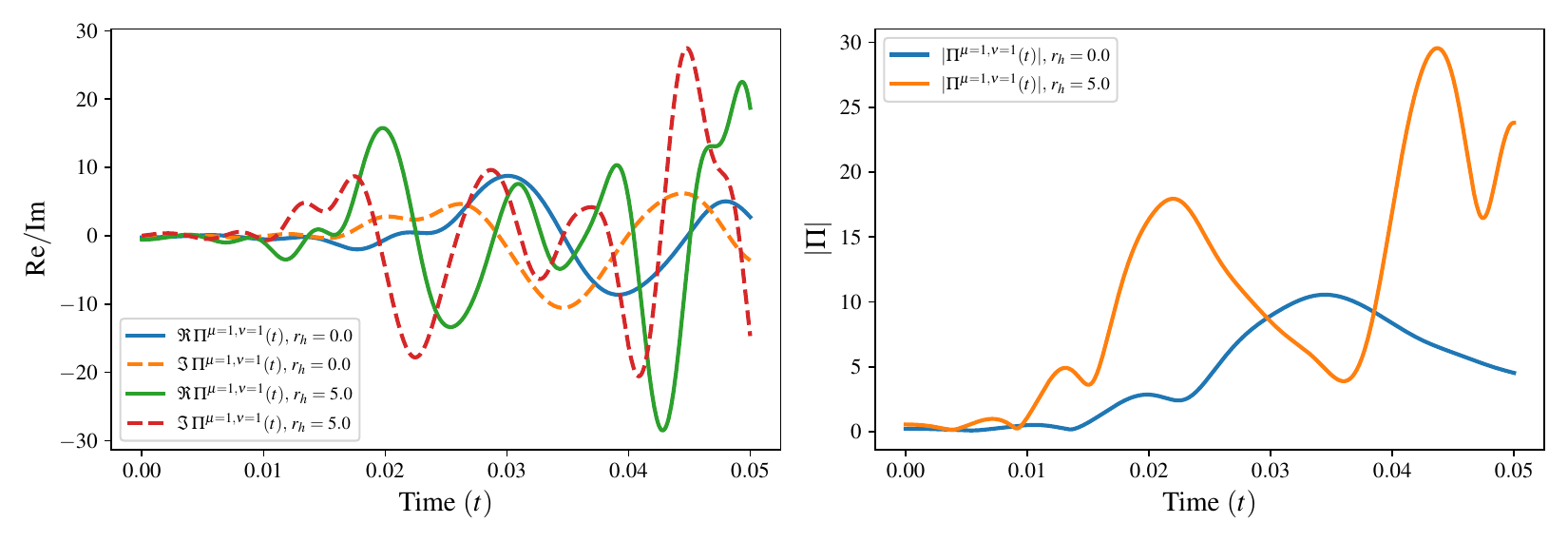}
\caption{
Real and imaginary parts of the charge-charge and current-current correlation functions $\Pi_{\mu\nu}(t,x_1,x_0)$ (left panels), and their absolute values (right panels), for the single-dipole quench with $m=1$, $N=40$, and separation $\Delta x = 10$. Results are shown for AdS$_2$ (blue) and AdS$_2$ black hole (orange) backgrounds, with horizon radius $r_h = 5$. Both correlators exhibit a negligible early time signal, followed by a pronounced peak when the right-moving front launched by the dipole reaches the probe site $x_1$, providing a time-of-flight diagnostic of the chiral wave. The current correlator $\Pi_{11}$ shows stronger oscillations and larger amplitude than the charge correlator $\Pi_{00}$, reflecting its enhanced sensitivity to the local couplings that encode the AdS redshift.
}
    \label{fig:corr}
\end{figure}

To explore this phenomenon, we fix $m=1$, $r_h=0$, $N=40$ and choose two sites separated by $\Delta x=10$, namely $x_0=15$ and $x_1=25$, to monitor the correlation between the anti\-fermion and fermion in the single-dipole quench. The resulting time traces $\Pi^{00}(t;x_1,x_0)$ and $\Pi^{11}(t; x_1,x_0)$ are shown in Fig.~\ref{fig:corr}. For the numerical study, we combine the charges on neighboring sites into a bond-resolved (physical) charge appropriate for the staggered-fermion discretization.

In both cases $(\mu,\nu)=(0,0)$ and $(1,1)$ there is a negligible early-time signal followed by a pronounced peak. The peak occurs at the time when the right-moving front launched by the dipole reaches the probe site $x_1$, i.e.\ when the wave in Figs.~\ref{fig:dipole_charge_and_EE} and \ref{fig:dipole_charge_and_EE_BH} (middle, $m=1$) arrives at $x=25$. In other words, $t_{\mathrm{peak}}\!\approx t_{\mathrm{front}}(x_0\!\to\!x_1)$, so the correlators provide a time-of-flight measurement of the chiral wave observed in the charge heatmap. This matches the light-cone picture established in Fig.~\ref{fig:LB}, where the fronts are confined within a LR cone.

While the timing of the main peak is set by the wavefront, the detailed lineshapes differ between charge and current. For $\Pi^{00}$ the real part rises to a single dominant peak, whereas for $\Pi^{11}$ stronger oscillations and a larger overall amplitude appear due to the derivative/current structure. In both cases the magnitude $|\Pi^{\mu\nu}|$ in Fig.~\ref{fig:corr} exhibits a single prominent lobe followed by damped oscillations, reflecting dephasing after the front passes the probe site.

\begin{figure}[H]
    \centering
    \includegraphics[width=\linewidth]{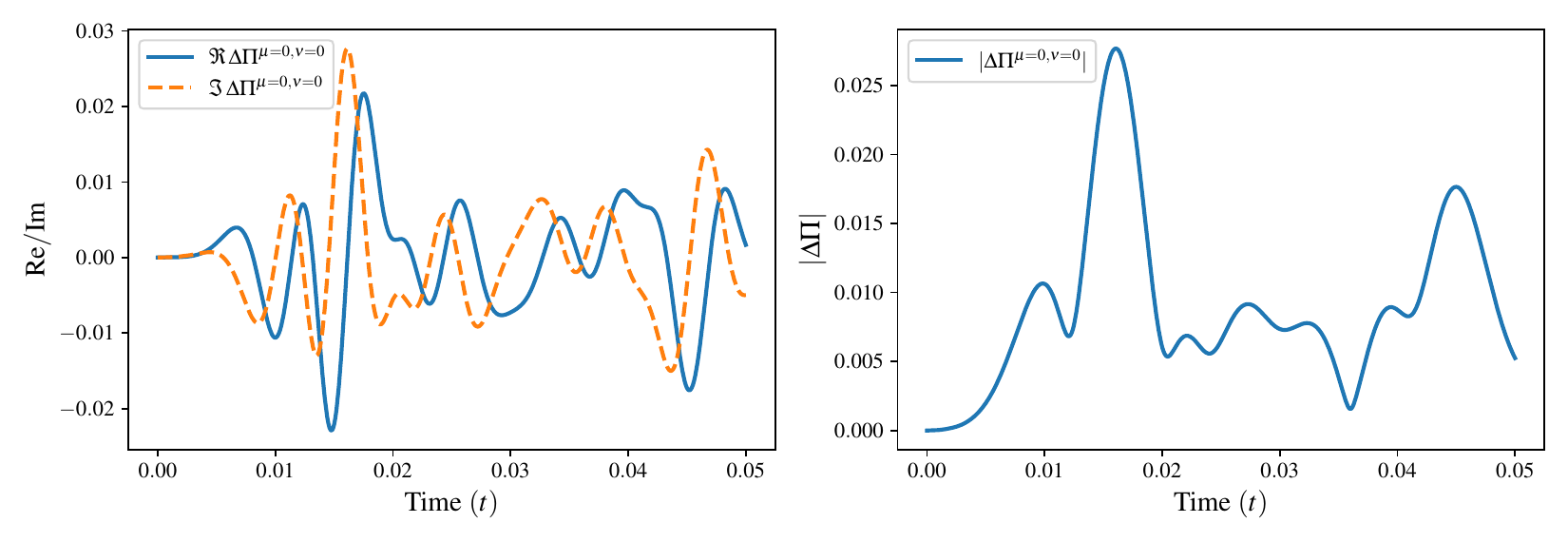}
    \includegraphics[width=\linewidth]{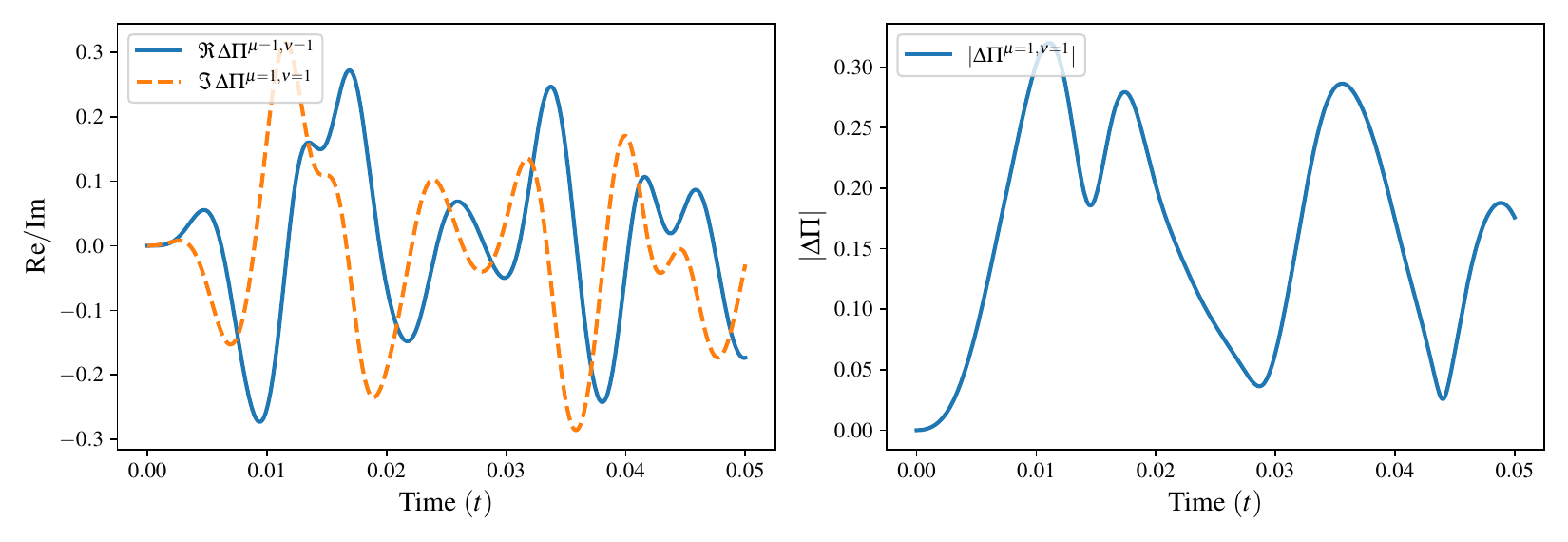}
\caption{
Real and imaginary parts of the orientation-odd difference 
$\Delta \Pi_{\mu\nu}(t; x_1, x_0) = \Pi_{\mu\nu}(t; x_1, x_0) - \Pi_{\mu\nu}(t; x_0, x_1)$ (left panels), and their absolute values (right panels), for the single-dipole quench with $m = 1$, $N = 40$, and separation $\Delta x = 10$. Results are shown for AdS$_2$ (blue), and AdS$_2$ blackhole (orange) geometries with horizon radius $r_h = 5$. The magnitude $|\Delta \Pi_{\mu\nu}|$ displays causality-limited peaks, whose timing coincides with the arrival of the chiral LR fronts at the probe site. The current correlator $\Delta \Pi_{11}$ exhibits a stronger asymmetry, and a larger amplitude than $\Delta \Pi_{00}$, highlighting its enhanced sensitivity to curvature-induced chiral transport.
}
    \label{fig:corr2}
\end{figure}

It is interesting that $\Pi^{\mu\nu}(t;x_1,x_0)$ is asymmetric ($\Pi^{\mu\nu}(t;x_1,x_0)\neq \Pi^{\mu\nu}(t;x_0,x_1)$) under the exchange $x_0\leftrightarrow x_1$, reflecting the underlying AdS geometry. Write the two-point functions as 
\begin{align}
\Pi^{\mu\mu}(t;x_1,x_0)\simeq \Pi^{\mu\mu}_{\mathrm{sym}}\!\left(t;|x_1-x_0|\right)+\Pi^{\mu\mu}_{\mathrm{asym}}(t;x_1,x_0). \label{eq:P00}
\end{align}
Here $\Pi^{\mu\mu}_{\mathrm{sym}}$ are orientation-even amplitudes controlled by the distance and the set of bonds between $x_0$ and $x_1$, and $\Pi^{\mu\mu}_{\mathrm{asym}}$ collects corrections due to spatial gradients of the position-dependent couplings and weights.

At the level of the difference
\begin{equation}
    \Delta\Pi^{\mu\nu}(t;x_1,x_0)\equiv \Pi^{\mu\nu}(t;x_1,x_0)-\Pi^{\mu\nu}(t;x_0,x_1)\;,
\end{equation}
we isolate the orientation-odd pieces $\Pi^{\mu\nu}_{\mathrm{asym}}(t;x_1,x_0)-\Pi^{\mu\nu}_{\mathrm{asym}}(t;x_0,x_1)$. It exhibits clear, causality–limited peaks whose timing is set by the chiral LR fronts. In Fig.~\ref{fig:corr2}, the first prominent maxima of $|\Delta\Pi^{00}|$ (top–right) and $|\Delta\Pi^{11}|$ (bottom–right) occur at the moment the right–moving front launched at $x_0$ first reaches $x_1$; the later lobes are post–front oscillations that remain strictly inside the LR cone. This identification follows directly from the LR overlays on the charge/entropy heatmaps (Fig.~\ref{fig:LB}), where the arrival–time curves predict the same front times that coincide with the leading peaks of $|\Delta\Pi^{\mu\nu}|$. Across channels, the current response amplifies the asymmetry. As confirmed in Fig.~\ref{fig:corr}, the peak magnitude in $|\Delta\Pi^{11}|$ is an order of magnitude larger than in $|\Delta\Pi^{00}|$, consistent with the stronger sensitivity of the current to the position–dependent couplings that set the LR slopes.

\section{\label{sec:scattering}Dipole-Dipole Collision}
In this section we study the real-time dynamics generated by preparing two localized dipole excitations at $t=0$ and evolving them under the curved-space Hamiltonian in AdS$_2$ and AdS$_2$ black hole backgrounds. It tracks the subsequent overlap and interference of the two curvature-modified causal fans. This setup cleanly exposes how the redshift and spin-connection terms control the geometry-dependent meeting time of the inward fronts and the onset of entanglement generation across the central cut.

Using the qubit Hamiltonian and matrix product state simulations, we observe the propagation and interference of the resulting charge and entanglement fronts. We show how curvature and horizons modify the causal structure of the collision process, producing left-right asymmetric light-cone dynamics. We relate the onset of central bipartite entanglement to the causal meeting of the inward LR branches, establishing a direct correspondence between entanglement growth and curvature-modified causality.

\subsection{\label{sec:scattering_charge}Collision of the Charges}

\begin{figure}
    \centering
    \includegraphics[width=0.32\linewidth]{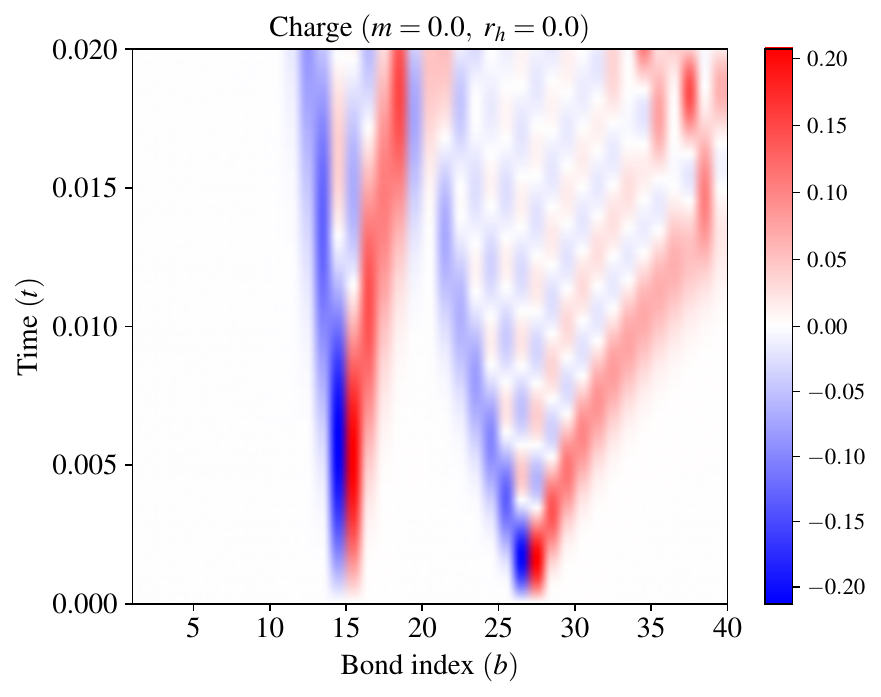}
    \includegraphics[width=0.32\linewidth]{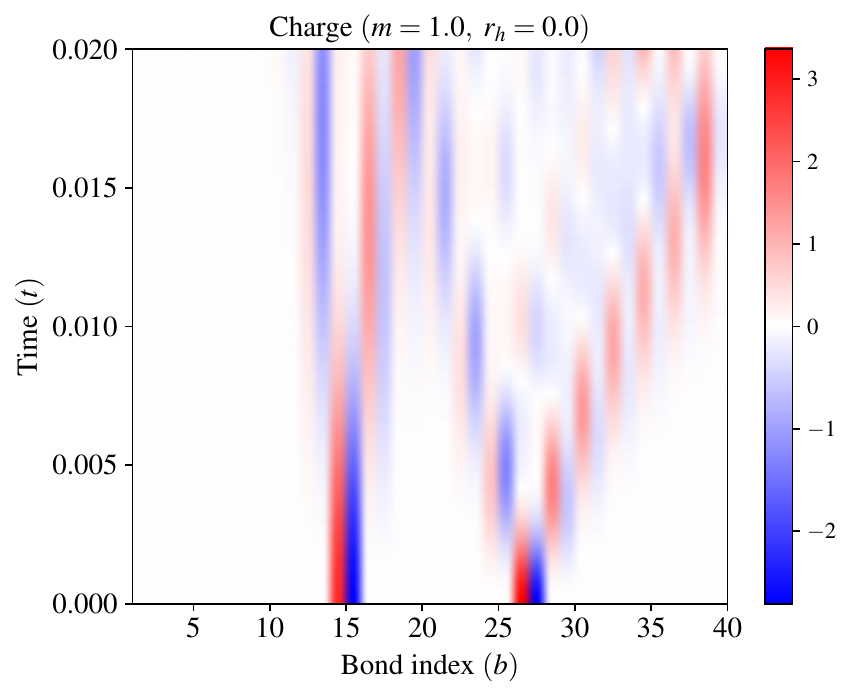}
    \includegraphics[width=0.32\linewidth]{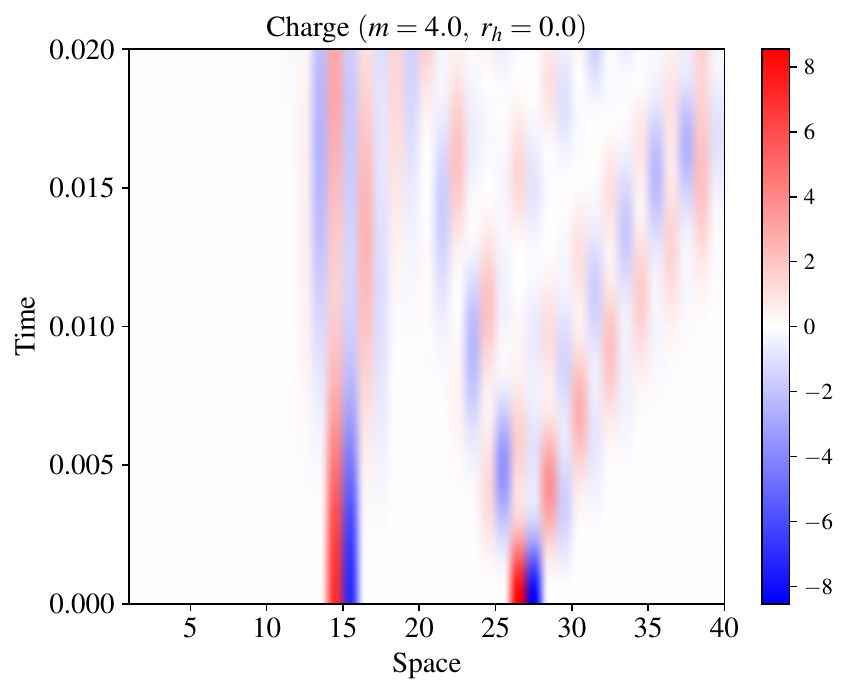}
    \includegraphics[width=0.32\linewidth]{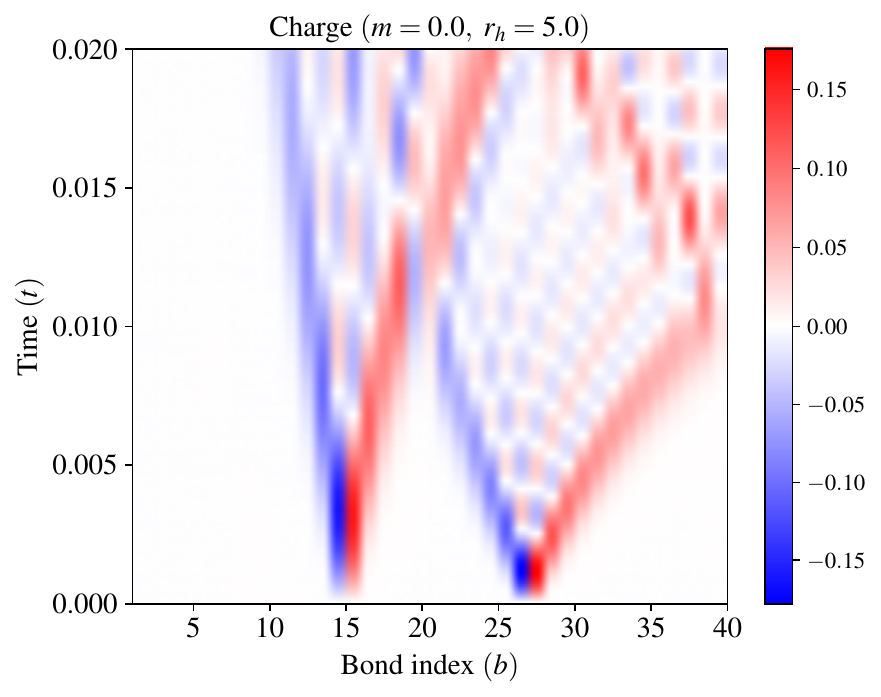}
    \includegraphics[width=0.32\linewidth]{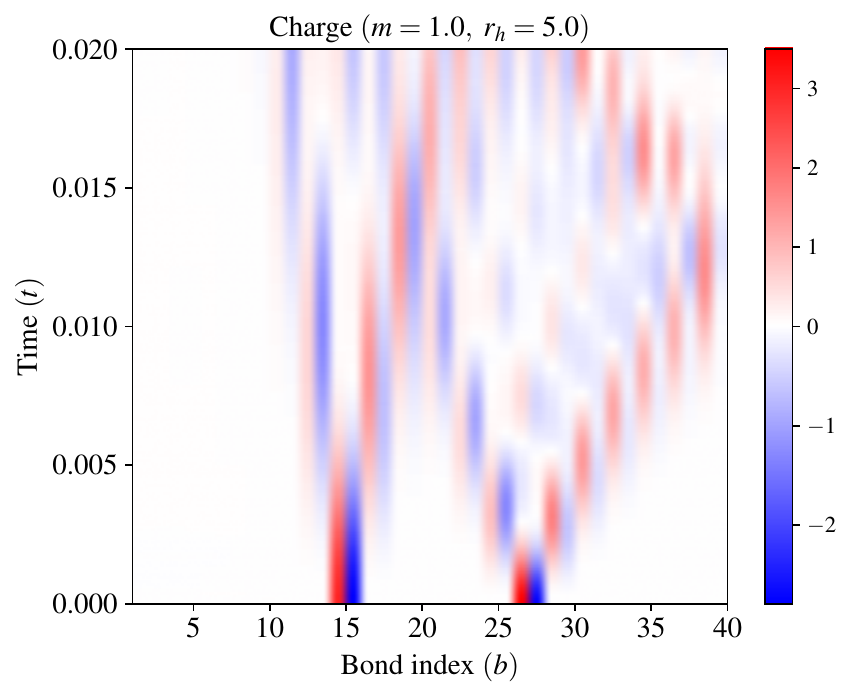}
    \includegraphics[width=0.32\linewidth]{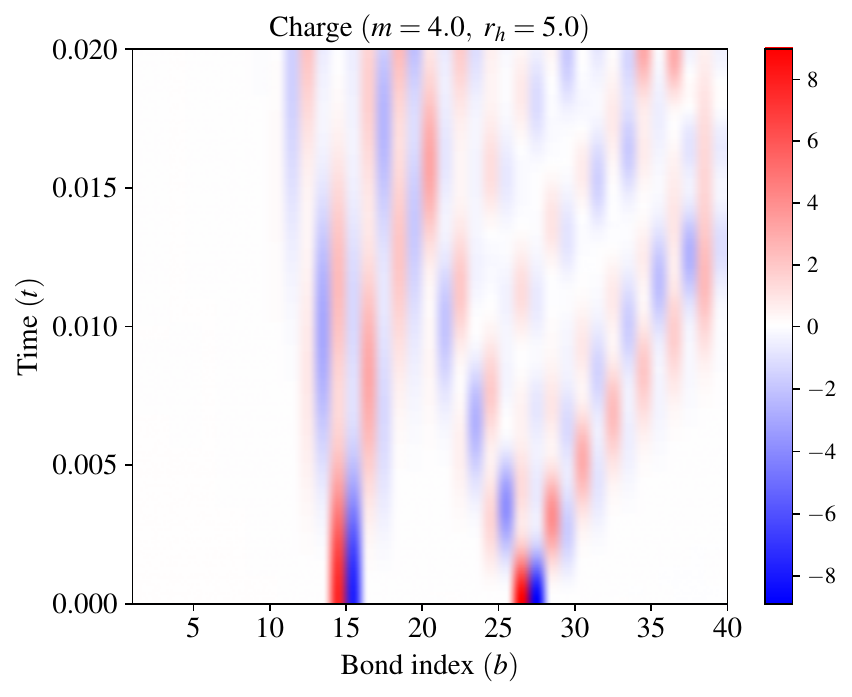}
\caption{
Real-time evolution of the local charge density during dipole-dipole collision in AdS$_2$ (top panels, $r_h = 0$), and AdS$_2$ black hole (bottom panels, $r_h = 5$) backgrounds. Each row shows results for fermion masses $m = 0$, $1$, and $4$ (from left to right), on a chain of $N = 40$ sites. Two initially separated dipoles launch left- and right-moving wavefronts, that form asymmetric causal fans, whose slopes encode the local redshift of the AdS geometry. When the inward branches meet near the center, the charge density focuses into a bright ridge, signaling the collision region. With $r_h\neq0$, a smaller $b$ brings it closer to the event horizon. Increasing $m$ or $r_h$ reduces the front velocities near the event horizon compared to those near the boundary, and suppresses post-collision oscillations, illustrating curvature and mass induced slowing of charge transport. 
}
    \label{fig:charge_scattering}
\end{figure}

Here we present the results, where the initial state is the ground state $\ket{\psi_{\mathrm{gs}}}$ of the Hamiltonian, obtained via MPS. We create two dipoles $\ket{\psi_{\mathrm{di-di}}}$ by flipping the spin-pairs in the ground state and evolve the system $\ket{\psi_{\mathrm{di-di}}(t)}=e^{-itH}\ket{\psi_{\mathrm{di-di}}}$ with the given Hamiltonian. Here $\ket{\psi_{\mathrm{di-di}}}$ is defined as:
\begin{align}
\ket{\psi_{\mathrm{di-di}}}\;=\;X_{\frac{N}{2}-\frac{d}{2}}X_{\frac{N}{2}-\frac{d}{2}+1}X_{\frac{N}{2}+\frac{d}{2}}X_{\frac{N}{2}+\frac{d}{2}+1}\ket{\psi_{\mathrm{gs}}}.
\end{align}
Then as eq.~\eqref{eq:subtract} we subtract the background propagation from the dipoles. 

The main purpose of the two-dipole protocol is to extract a genuinely geometry-dependent real-time scale: the time at which the \emph{inward} causal branches meet and can begin to generate entanglement across the central cut. In a homogeneous chain this meeting time is essentially $t\sim d/(2v)$, while in AdS$_2$/AdS$_2$ black hole it depends on the position-dependent couplings set by the redshift and spin connection. The inhomogeneous LR arrival-time construction provides a parameter-free prediction for this meeting time, which we test quantitatively against the onset time of the central bipartite entropy in Fig.~\ref{fig:tE_vs_tLR}.

Fig.~\ref{fig:charge_scattering} shows the results for AdS space.  Two causal cones emanate from the initially separated dipoles at sites $N/2\pm d/2$. Each fan is the counterpart of the single–dipole chiral wave of Fig.~\ref{fig:dipole_charge_and_EE}, where a left– and a right–moving branch whose slopes encode the inhomogeneous, position–dependent propagation speed in AdS. As a result, the pattern is manifestly chiral—the inward (toward the center) and outward branches have different slopes—reflecting the local redshift of the AdS geometry. When the two inward branches meet, the charge signal focuses at the collision region and produces a bright ridge. The current panels simultaneously exhibit strong sign–alternating streaks that emphasize the directionality of transport. Lighter fermions ($m=0$) produce broader, faster cones with pronounced interference fringes, while heavier fermions ($m=4$) show slower fronts, reduced penetration, and suppressed oscillations.
Increasing the black hole radius $r_h$ further reduces the slopes of the fans and damps post–collision echoes, consistent with the redshift–induced slowdown.

\begin{figure}
    \centering
    \includegraphics[width=0.49\linewidth]{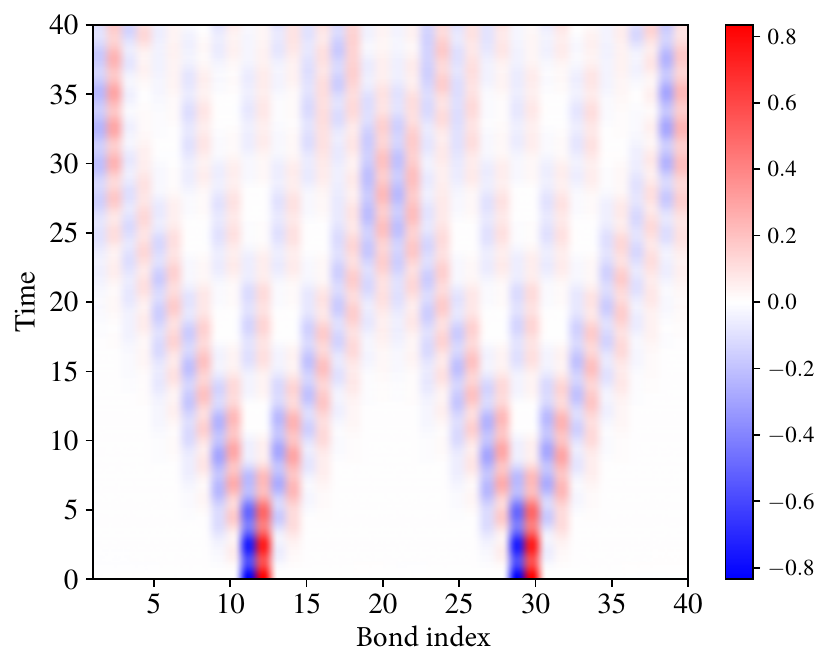}
    \includegraphics[width=0.49\linewidth]{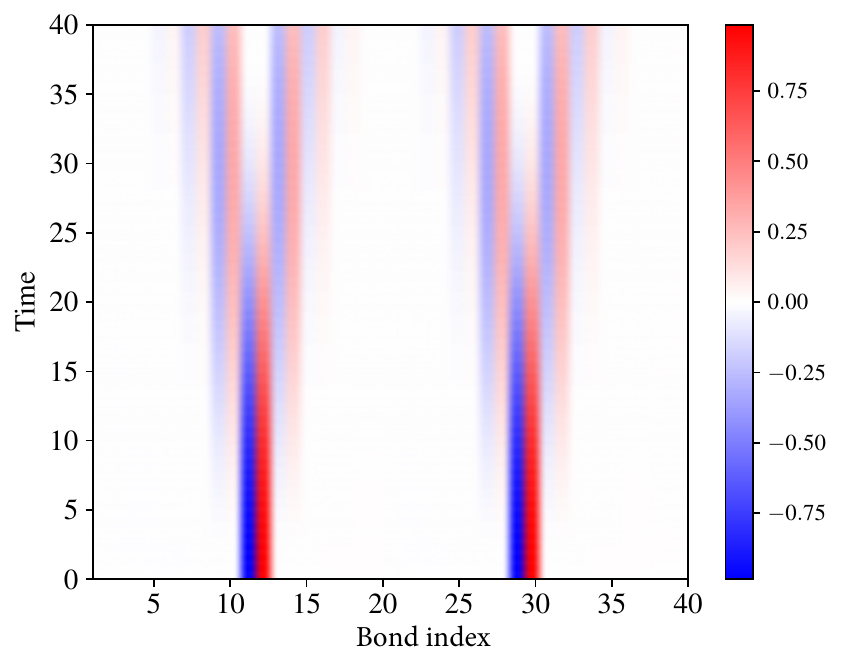}
\caption{
Real-time evolution of the local charge density of two dipoles in flat space for fermion masses $m = 1$ (left) and $m = 4$ (right). In the absence of curvature ($\alpha_n \to 1$), the propagation becomes parity-symmetric: the two light-cone fronts are straight, and exhibit equal left and right velocities. Increasing the fermion mass uniformly reduces the propagation speed, while preserving this symmetry. 
These results contrast with the AdS$_2$ and AdS$_2$ black hole cases, where redshift and spin connection effects break left-right symmetry, and slow the fronts near the horizon.
}
\label{fig:flat_charge_scattering}
\end{figure}

In contrast, in flat space (Fig.~\ref{fig:flat_charge_scattering}), the hopping $h_n$ become homogeneous and renders the LR speed uniform\footnote{The flat space limit can be obtained by taking $\alpha_n\to 1~(\forall n),~L\to\infty$ in the Hamiltonian \eqref{eq:Ham_AdS}.}. Therefore, the two cones are straight and parity‑symmetric, $t^{\rm L}_{\rm front}=t^{\rm R}_{\rm front}$, and masses primarily set a common time scale, by uniformly reducing the propagation velocity, while preserving the left–right symmetry of the patterns. These trends are consistent with the chiral propagation seen in the single‑dipole charge/entropy heatmaps, and with the inhomogeneous LR bounds, which confine all measured fronts inside the causal cone in AdS/AdS black hole, and collapse to symmetric, straight cones in the flat space limit.

In the flat case, $v_{\rm loc}$ is spatially uniform, and $t_{\rm flat}\simeq \Delta x/(\kappa/(4a))$ is linear in the separation. In AdS/AdS black hole, the redshift $\alpha_n$ suppresses $J_n$ near the horizon ($\alpha_n\!\to\!0$) as $r_n\!\to\!r_h$ and the spin-connection term $D_n$ produces an additional chiral contribution to $\|h_{n,n+1}\|$ \eqref{eq:hopping_term}, which gives $v(n)=2\sqrt{J_n^2+D_n^2}$. In the flat case, $J_n=\tfrac{1}{4a},D_n=0$ and in the AdS case, $J_n=\tfrac{\alpha^2_n}{4a},D_n=\tfrac{an}{8L^2}$. Consequently, segments of the path that pass through small $\alpha_n$ regions have reduced $v_{\rm loc}$, and dominate the sum $\sum 1/v_{\rm loc}$, yielding longer arrival times than in flat space:
\begin{equation}
\frac{t_{\rm AdS/BH}}{t_{\rm flat}}
\;\sim\;\frac{1}{\Delta x}\sum_{\ell\in{\rm path}}
\frac{1}{\sqrt{\alpha_\ell^{4}+\tfrac{a^2 n_\ell^2}{4L^4}}}\;>\;1\;.
\end{equation}
Thus, relative to Fig.~\ref{fig:flat_charge_scattering}, the wavefronts in Fig.~\ref{fig:charge_scattering} unfold on a longer (stretched) time scale set by the spatially varying $J_n$ and $D_n$. Paths that approach the horizon (smaller $\alpha_n$) are the slowest, while those toward larger $\alpha_n$ are faster. When $r_h \neq 0$, decreasing $b$ brings it nearer to the event horizon of the black hole. As a result, waves with smaller $b$ values travel more slowly compared to those with larger $b$ values.

Particle collisions in different (1+1)-dimensional flat space models have been studied, for example
in \cite{Farrell:2025nkx,qr72-51v1,Chai:2023qpq,Barata:2025rjb}.

\subsection{\label{sec:scattering_LR}Entanglement Growth vs Lieb–Robinson Causality}
\begin{figure}
    \centering
    \includegraphics[width=0.49\linewidth]{AdS_BH_two_dipoles_ent_line_plot_MPS_gs_evolv_N40_r00.0.pdf}
    \includegraphics[width=0.49\linewidth]{AdS_BH_two_dipoles_ent_line_plot_MPS_gs_evolv_N40_r05.0.pdf}
    \includegraphics[width=0.49\linewidth]{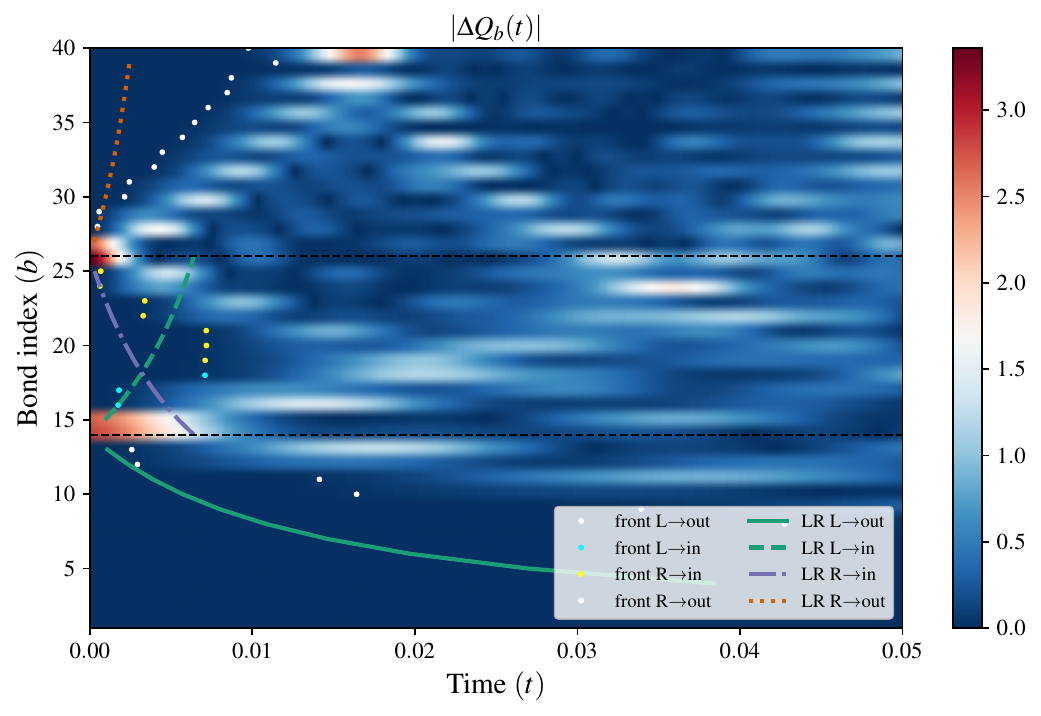}
    \includegraphics[width=0.49\linewidth]{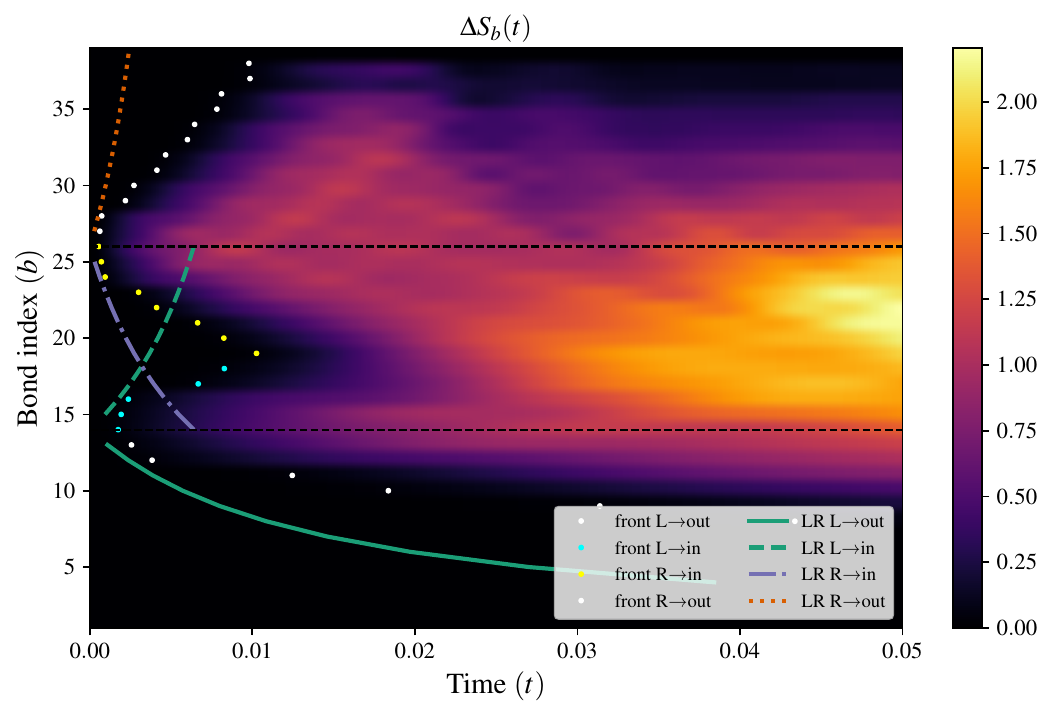}
    \includegraphics[width=0.49\linewidth]{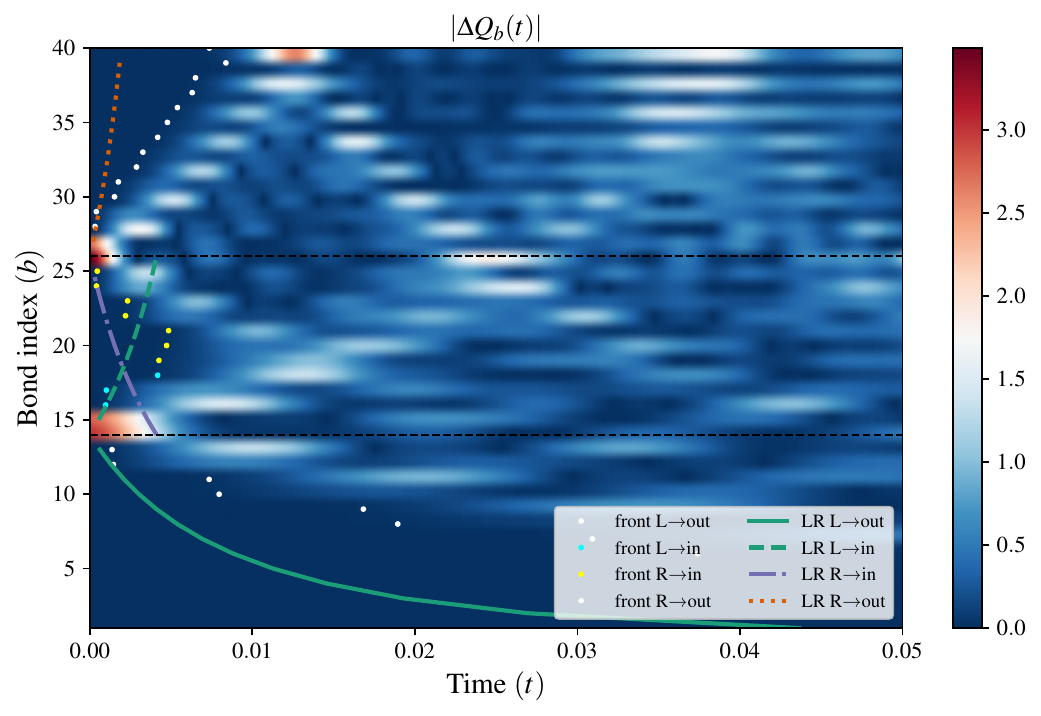}
    \includegraphics[width=0.49\linewidth]{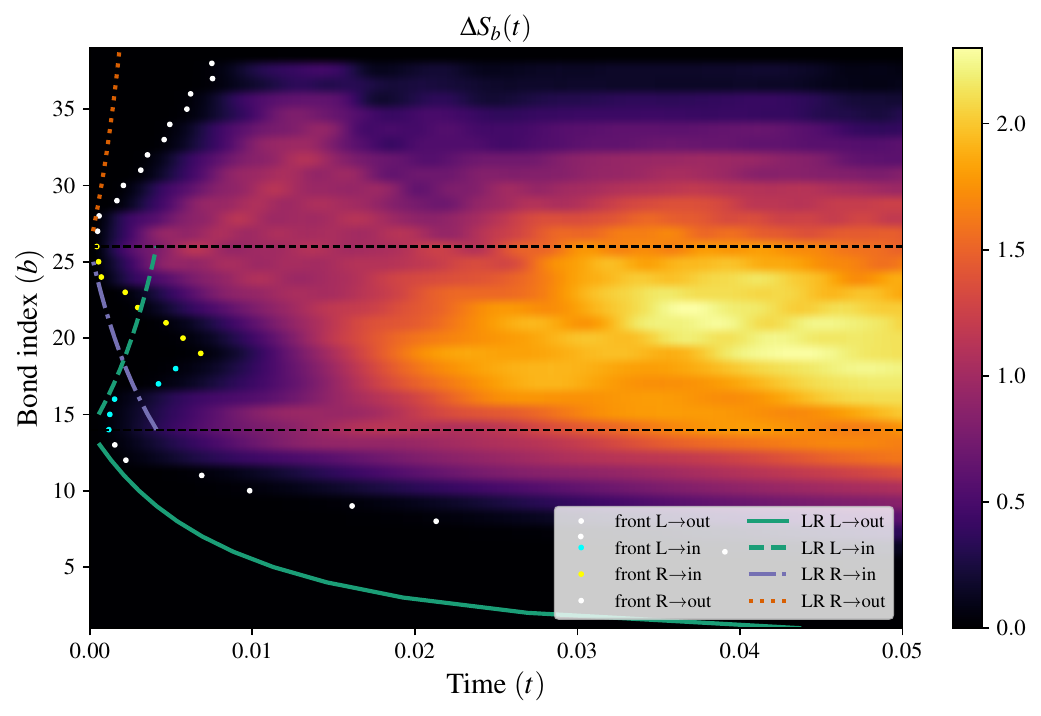}
    \caption{Real-time evolution of charge and entanglement entropy in the setting of dipole-dipole collision in AdS$_2$ and AdS$_2$ black hole backgrounds. Top panels show the time-dependence of the bipartite entanglement entropy ($r_h=0$ (left) and $r_h=5$ (right)). The dashed-lines in the heatmaps correspond to the initial positions of the dipoles, which are separated by a distance $d=10$. In the middle $(r_h=0)$ and bottom $(r_h=5)$ panels, two causal fans of charge $|\Delta Q_b(t)|$ and entanglement entropy  $\Delta S_b(t)$ launch from the dipoles and interfere near their meeting region, with left–right asymmetry due to the AdS redshift. All fronts inside the inhomogeneous LR cone. Larger $r_h$ reduces front speeds and amplitudes near the event horizon compared to those near the boundary. When $r_h \neq 0$, decreasing $b$ moves it nearer to the event horizon. At later times, one of the waves encounters the boundary and begins to travel back toward the center. ($N=40,m=1$)}
    \label{fig:two_source_LR}
\end{figure}

Fig.~\ref{fig:two_source_LR} shows the entanglement dynamics during dipole–dipole collision ($N=40$). The top panels show the bipartite entanglement across the central cut, $\Delta S_{L|R}(t)$. In contrast to the single‑dipole case (Fig.~\ref{fig:EE}), $\Delta S_{L|R}(t)$ exhibits a short early‑time plateau before the mutual light cones overlap. As the dipoles approach and begin to exchange correlations, the entropy grows steadily. The bottom panels display the local entanglement $\Delta S_b(t)$ for $m=1$ and $r_h=5$. Two light‑cone fronts emanate from the initially separated dipoles, and a bright ridge appears where their trajectories intersect. Outside these cones the signal remains negligible, consistent with LR causality. 

\if{
\begin{figure}[H]
    \centering
    \includegraphics[width=0.49\linewidth]{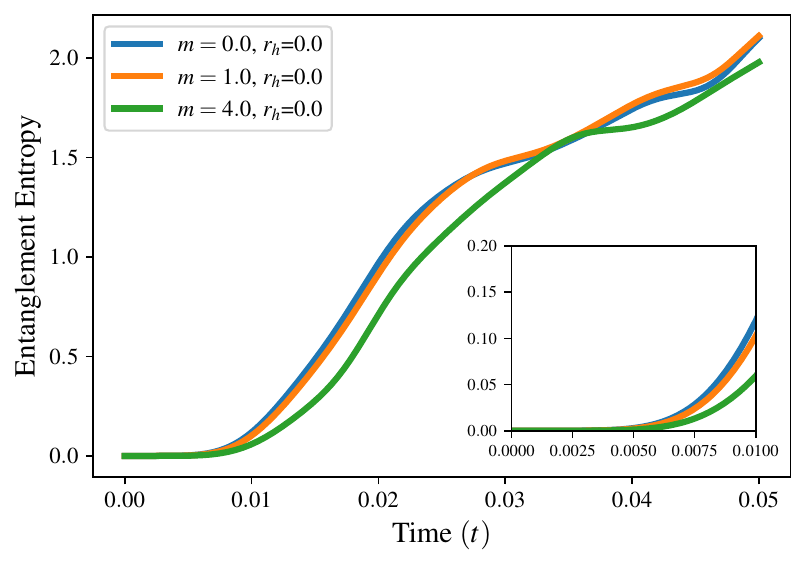}
    \includegraphics[width=0.49\linewidth]{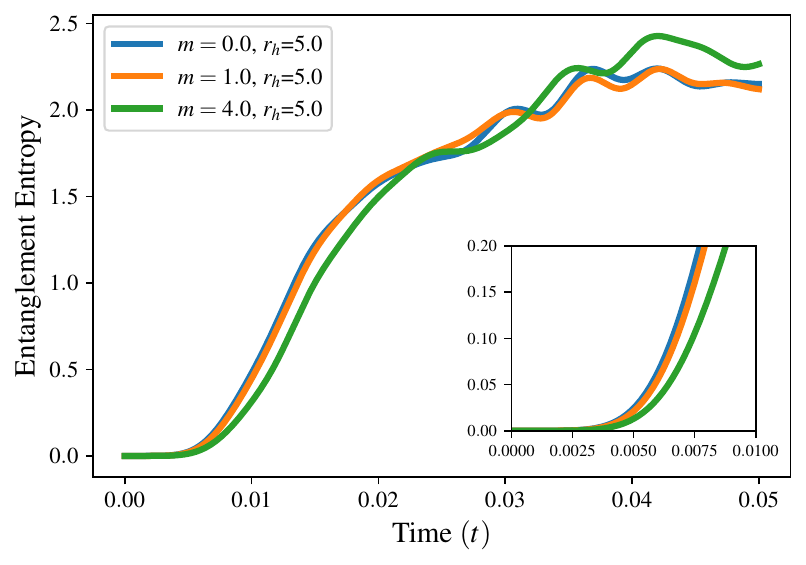}
    \includegraphics[width=0.49\linewidth]{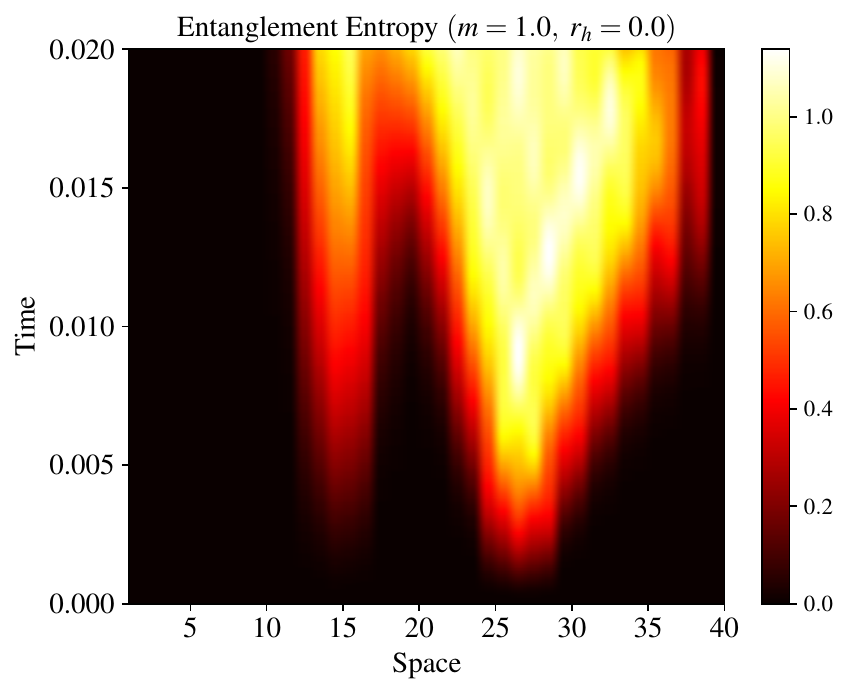}
    \includegraphics[width=0.49\linewidth]{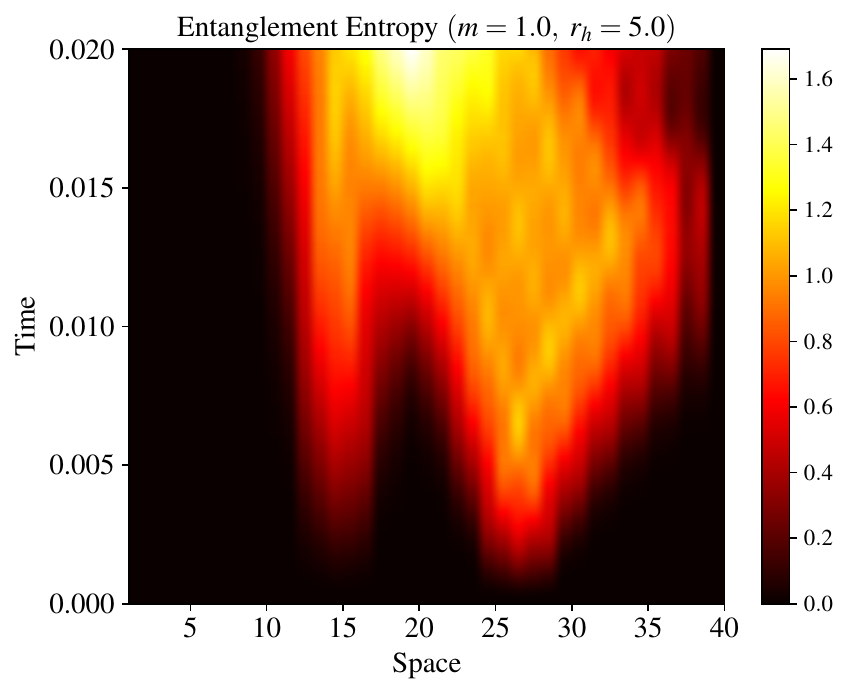}
    \caption{Time-evolution of the entanglement entropy $(N=40)$. Top panels show the time-dependence of the bipartite entanglement entropy and the bottom panels show the local entanglement entropy. From left to right: $r_h=0,5$. }
    \label{fig:EE_scattering}
\end{figure}
}\fi

The early‐time plateau of the central bipartite entropy $\Delta S_{L|R}(t)$ (the small windows in the top panes of Fig.~\ref{fig:two_source_LR}) is naturally explained by the two–source LR analysis in Fig.~\ref{fig:two_source_LR} (middle and bottom panels): the central cut does not gain long‐range correlations until the inward LR branches from the left and right sources (L$\!\to$in, R$\!\to$in) arrive at the cut. The subsequent growth of $\Delta S_{L|R}(t)$ coincides with the overlap region where the two inward fronts intersect, which also appears as the bright ridge in the local entanglement heatmap $\Delta S_b(t)$. Outside the cones the signal remains negligible, consistent with LR causality. At later times, as the outward branches (L$\!\to$out, R$\!\to$out) carry excitations away and screening sets in, the rate of entanglement production diminishes and $\Delta S_{L|R}(t)$ approaches a saturation regime. The left–right asymmetry of the four LR branches (different slopes for L/R, in/out) quantifies the chiral propagation in AdS and explains the asymmetric growth rates.

\begin{figure}
    \centering
    \includegraphics[width=0.5\linewidth]{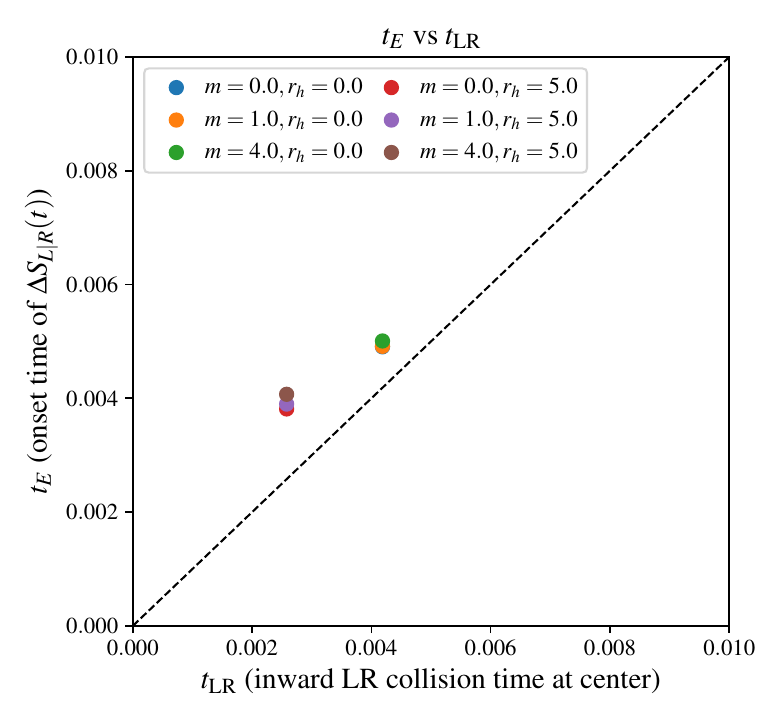}
    \caption{Onset time $t_E$ of entanglement vs LR collision time $t_{LR}$. For each $(m,r_h)$ we compare the entanglement onset $t_E$ from the central entropy trace to the inhomogeneous LR collision time $t_{\mathrm{LR}}$ of the two inward branches at the central bond. Points lie close to $y=x$, showing that the central entanglement begins when the inward LR cones meet. ($L=1$, $N=40$)}
    \label{fig:tE_vs_tLR}
\end{figure}

For the simulation of dipole collision, we take the source bonds $b_L=\frac{N}{2}-\frac{d}{2}-1$ and $b_R=\frac{N}{2}+\frac{d}{2}+1$, and the central bond $b_c=\frac{N}{2}$. The inhomogeneous LR arrival time for the two inward branches to reach the center is
\begin{equation}
t_{\mathrm{LR}}
=\max\left\{
\sum_{\ell=b_L}^{b_c-1}\!\frac{1}{v_{\mathrm{loc}}(\ell)}\;,\;
\sum_{\ell=b_c}^{b_R-1}\!\frac{1}{v_{\mathrm{loc}}(\ell)}\right\}.
\end{equation}
To extract the entanglement onset we analyze the central bipartite entropy
$\Delta S_{L|R}(t)=S_{L|R}(t)-S_{L|R}(0)$ and define $t_E$ as the earliest sustained rise after the early–time plateau (see Fig.~\ref{fig:two_source_LR}). We smooth $\Delta S_{L|R}(t)$, estimate a robust noise floor from the pre‑collision segment, require both amplitude and slope to exceed that floor for several consecutive steps. Across $(m,r_h)\in\{0,1,4\}\times\{0,5\}$ the scatter of $t_E$ vs $t_{\mathrm{LR}}$ clusters near the $y=x$ line (Fig.~\ref{fig:tE_vs_tLR}), demonstrating that the central entanglement begins precisely when the two inward LR branches meet at the center. This agrees with the line plots, where the onset occurs near $t\simeq0.005$ for $r_h=0$ and $t\simeq0.003$ for $r_h=5$, and with the LR‑overlay heatmaps where the $L\!\to\!\mathrm{in}$ and $R\!\to\!\mathrm{in}$ curves intersect at the same times. If $r_h \neq 0$, the closer $b$ is to 0, the closer it is to the black hole horizon. Therefore, a wave with a smaller $b$ propagates more slowly than one with a larger $b$.

\section{\label{sec:conclusion}Conclusion and Discussion}
We simulate the real-time chiral gravitational dynamics of Dirac fermions in AdS$_2$ and AdS$_2$ blackhole backgrounds, which makes the geometric origin of chiral transport explicit. In this framework, the spin connection arising from the geometry acts as a position-dependent chiral chemical potential, manifested through the redshifted couplings in the lattice Hamiltonian. These features give rise to robust, left–right asymmetric, and nonlinear wavefronts, whose propagation speeds decrease with increasing fermion mass $m$ and horizon radius $r_h$. Entanglement production is restricted within an inhomogeneous LR cone, and displays distinct chirality: the initial growth of local or bipartite entropy follows different left/right slopes, directly reflecting the spatially varying redshift.

When two dipoles propagate toward one another, two asymmetric causal cones emanate from the sources, and interfere near their meeting region. The central bipartite entropy displays an early‑time plateau, that lifts precisely when the inward LR branches collide at the central cut. The same meeting point appears as a bright ridge in the local entropy heatmap. A direct comparison of the measured onset time with the LR collision time shows near one‑to‑one agreement across $(m,r_h)$, confirming that entanglement growth at the center is triggered by the causal arrival of the inward fronts. Increasing $r_h$ slows all branches, and damps post‑collision echoes, highlighting the role of horizons as efficient suppressors of propagation and entanglement. In the flat limit the redshift becomes uniform, and the left–right asymmetry disappears. This provides a stringent check that the observed chiral effects are geometric in origin.

The framework developed here suggests several immediate avenues for quantum computing and high energy theory. A natural next step is to couple the AdS and dS Dirac matter to a dynamical gauge field, and study how curvature–induced chirality interferes with confinement dynamics and the quantum anomaly in curved spacetime QED/QCD. The correlators/operators and inhomogeneous LR bounds diagnostics, that proved effective here, can serve as quantitative probes, enabling a clean comparison between curved and flat backgrounds within a unified real–time protocol. It is also compelling to explore beyond $(1+1)$d, and to test robustness in settings with interactions, disorder, or long‑range couplings, as well as to incorporate effective backreaction, by allowing the local couplings to respond to energy density. Multi‑source protocols, generalizing the dipole–dipole setup, could map out interference, screening, and entanglement hydrodynamics in curved backgrounds, while larger system calculations would sharpen the scaling of the inhomogeneous LR bounds. 

The qubit Hamiltonian \eqref{eq:Ham_AdS} is a nearest-neighbor $XY$ chain with a bond-dependent DM interaction (equivalently, complex hopping) and a staggered on-site $Z$ field. Such interactions are natural targets for programmable platforms that support tunable nearest-neighbor exchange with controllable phases, as well as for digital simulation.

Digitally, one may split $H$ into commuting on-site terms and even/odd bond layers,
$H=H_Z+H_{\rm even}+H_{\rm odd}$, and apply a product-formula (Trotter) evolution. Each
two-site bond unitary $\exp[-i\Delta t(J_n(XX+YY)+D_n(XY-YX))]$ can be implemented with a constant number of entangling gates plus single-qubit rotations, with angles set by $(J_n,D_n)\Delta t$. The resulting two-qubit gate count per Trotter step scales as $\mathcal{O}(N)$ (with an even/odd pattern over $N-1$ bonds), while single-qubit $Z$ rotations implement the staggered mass term. A detailed hardware-specific compilation is beyond our present scope, but the mapping is explicit and can be used directly for resource estimates once a target platform and gate set are fixed.

\section*{Acknowledgement}
This work is supported by the NSF under Grant No. OSI-2328774 (KI), by the Israeli Science Foundation Excellence Center, the US-Israel Binational Science Foundation, the Israel Ministry of Science (YO).

\bibliographystyle{JHEP}
\bibliography{ref}

\end{document}